\documentclass[reprint,amsmath,amssymb,aps,longbibliography,floatfix,raggedbottom,superscriptaddress]{revtex4-2}
\usepackage[T1]{fontenc}
\usepackage{lmodern}
\usepackage{amsmath,amssymb,bm,graphicx,dcolumn,booktabs}
\usepackage[caption=false]{subfig}
\usepackage{capt-of}
\usepackage{xcolor}
\usepackage[colorlinks=true,linkcolor=blue,citecolor=blue,urlcolor=blue]{hyperref}
\newcommand{\ii}{\mathrm{i}}
\newcommand{\dd}{\mathrm{d}}
\newcommand{\EP}{\mathrm{EP}}
\newcommand{\ch}{\widehat\chi}
\newcommand{\Ree}{\operatorname{Re}}
\newcommand{\Imm}{\operatorname{Im}}
\newcommand{\norm}[1]{\left\lVert #1\right\rVert}

\newcommand{\CC}{\mathcal C}
\newcommand{\SSS}{\mathcal S}
\graphicspath{{figures/}}

\begin{document}
\title{Identifying Exceptional Points in Black Hole Ringdowns}

\author{Zhen-Xiao Zhang}
\email{zx.zhang@mail.nankai.edu.cn}
\affiliation{School of Physics, Nankai University, 94 Weijin Road, Tianjin 300071, China}

\author{Chen Lan}
\email{stlanchen@126.com}
\affiliation{Department of Physics, Yantai University, 30 Qingquan Road, Yantai 264005, China}

\author{Yi Wang}
\email{phyw@ust.hk}
\affiliation{Department of Physics, The Hong Kong University of Science and Technology, Clear Water Bay, Kowloon, Hong Kong, P.R. China}
\affiliation{The HKUST Jockey Club Institute for Advanced Study, The Hong Kong University of Science and Technology, Clear Water Bay, Kowloon, Hong Kong, P.R. China}

\begin{abstract}
Ringdown spectroscopy probes black hole (BH) dynamics, and exceptional points (EPs) can reveal underlying degrees of freedom. We develop a procedure to identify EPs in ringdowns using a time-weighted Jordan term and a splitting bound. For relative Jordan excitation proportional to separation, we derive a sixth-order squared distance between split and EP waveforms after fitting the resonance center. Independent constraints on these quantities can restore fourth order. Complete waveforms show that omitting ordinary responses can strengthen EP evidence and spuriously tighten splitting bounds.
\end{abstract}
\maketitle
\setcounter{dbltopnumber}{1}
\raggedbottom
\section{Introduction}
\label{sec:introduction}

As a perturbed BH settles down, the emitted radiation carries information about its dynamical degrees of freedom.
During ringdown, quasinormal modes describe oscillations that decay at characteristic rates \cite{BertiCardosoStarinets2009}; measuring several modes tests whether they belong to the same remnant spacetime \cite{Isi2019,LVK2026GW250114}. Radiation escaping to infinity and entering the horizon makes this spectral problem non-Hermitian. Nearly equal frequencies and damping rates can accompany different dynamics. Two modes may remain independent, or their spatial profiles may also merge, leaving only one independent mode. 
Such coalescence defines a second-order exceptional point (EP), distinct from an ordinary degeneracy~\cite{Kato1966,MiriAlu2019}, and changes the collective time dependence of the radiation. Distinguishing coalescence from a split pair would reveal how these degrees of freedom participate in BH relaxation.

EPs have been found in scalar perturbations of BHs \cite{Cavalcante2024,Cheng2026},
and split pairs can exhibit enhanced excitation near avoided crossings in gravitational and coupled-field systems \cite{Motohashi2025,TakahashiMotohashiTakahashi2025}. Collective descriptions relate mode coalescence to a damped oscillation multiplied by time, called the Jordan term \cite{YangBertiFranchini2025,Macedo2025,MorikawaOgawaHirose2026}. Harmonic inversion can extract polynomial EP responses \cite{Fuchs2014}, and a uniform theory describes non-Hermitian response near and at EPs \cite{BidSchomerus2024}. Detectability has also been studied when individual frequencies remain unresolved \cite{ImafukuOshitaTakeda2026}. Yet a nearby split pair can reproduce the same leading time dependence over a finite observation. Jordan evidence must therefore be accompanied by a tight splitting region allowed by the data and containing zero. The largest mode separation it allows specifies the resolution of an observational coalescence claim.

This bound depends on how the measured waveform is decomposed. Other damped components outside the tested pair, termed ordinary responses, interfere coherently with it \cite{MaCleaning2023,DyerMoore2026}. Omitting such radiation can strengthen Jordan evidence and spuriously tighten the splitting bound. 
Including these ordinary responses in the fit controls the bias, but letting their amplitudes and frequencies vary freely also allows the ordinary-response model to absorb differences between coalesced and split responses.
Reliable modeling and attainable resolution are thus linked: a convincing identification must survive the freedom needed to describe the ordinary responses.

We develop a procedure for identifying EPs in gravitational-wave waveforms, accounting for finite observations, ordinary responses, noise, and polarization readout. After defining the joint criterion, we derive analytically its excitation-dependent resolution. When relative Jordan excitation (the time-weighted contribution relative to the unweighted oscillation over one decay time) decreases linearly with mode separation, fitting the resonance center---the mean frequency and decay rate---absorbs the leading waveform difference caused by splitting. The squared distance between the split waveform and the best-fitting EP waveform then scales as the sixth power of the mode separation measured relative to the decay rate, leaving a very small difference to resolve near coalescence.
Complete BH waveforms and simulated strain then 
test how omitting ordinary responses biases the fit, and how much resolution is lost when additional ordinary-response parameters are fitted.
Finally, we show how independent information constraining the center can prevent this compensation and retain a fourth-power contribution to the squared distance, making small splittings easier to resolve.

\section{Exceptional-point response and identification}
\label{sec:response}

An individual mode away from an EP produces a damped oscillation. We call the two-mode contribution being tested the EP cluster (for brevity and clarity, this term covers both a closely spaced pair of complex frequencies and an actual EP where the modes coalesce). At an EP, its response takes the form
\begin{equation}
    H_P^{\EP}(\tau)=e^{(-\gamma_c+\ii\Omega_c)\tau}(C_P+J_P\tau).
    \label{eq:ep_response}
\end{equation}
Here $\tau$ is time from a common reference, $\Omega_c$ is the oscillation frequency, and $\gamma_c>0$ is the decay rate. The channel label $P$ can denote a radiation polarization or a separately excited response. The complex coefficients $C_P$ and $J_P$ specify the magnitude and phase of the constant and time-linear contributions; the time-weighted exponential contribution is the Jordan term. Their values depend on the perturbation and on how the radiation is read out. An EP can therefore have a weak or absent measured Jordan term.

For a nearby split pair, write the two complex exponents as $s_\pm=s_c\pm\sqrt{\chi}$, where $s_c=-\gamma_c+\ii\Omega_c$ is their mean. The complex parameter $\chi$ describes splitting in frequency and decay rate, with $\chi=0$ at coalescence; we use its dimensionless form $\ch=\chi/\gamma_c^2$. On a fixed finite time interval, the EP cluster's response expands as
\begin{equation}
    \begin{aligned}
        H_P(\tau)=e^{s_c\tau}\bigg[&C_P+J_P\tau\\
        &+\chi\left(\frac{C_P\tau^2}{2}+\frac{J_P\tau^3}{6}\right)
        +O(\chi^2)\bigg].
    \end{aligned}
    \label{eq:response}
\end{equation}
Both EP and split responses contain the constant and linear terms. Resolving splitting requires detecting the smaller, higher-power time dependence that follows them. We fit the exact response, derived in Supplemental Material (SM), Sec.~I \cite{SupplementalMaterial}; Eq.~\eqref{eq:response} exposes the difference that the data must resolve.

The measured waveform combines the real parts of $H_P$ with channel-dependent response factors, damped components outside the EP cluster, termed ordinary responses, late-time tails, and noise. For the EP cluster, we fit a single-exponential model ($J=\chi=0$), an EP model ($\chi=0$), and a split model (complex $\chi$ free, including zero), keeping the same freedom in the ordinary responses and tails. Each fit optimizes its center and all allowed amplitudes and ordinary frequencies. If $\ell_1$, $\ell_{\EP}$, and $\ell_{\rm split}$ denote the respective maximum log likelihoods with the specified noise covariance \cite{IsiFarr2021}, then
\begin{equation}
    T_J=2(\ell_{\EP}-\ell_1),\qquad
    T_\chi=2(\ell_{\rm split}-\ell_{\EP}).
    \label{eq:statistics}
\end{equation}
Thus $T_J$ (Jordan evidence) measures the improvement from allowing the Jordan term, and $T_\chi$ (splitting evidence) the improvement from allowing nonzero splitting.

At a prescribed resolution, we identify coalescence by requiring evidence for the Jordan term, compatibility with zero splitting, and a sufficiently small upper bound on the allowed separation:
\begin{equation}
    T_J>t_J,\qquad T_\chi\le t_\chi,\qquad
    \chi_{\rm up}+b_\chi<\chi_*.
    \label{eq:criterion}
\end{equation}
Here $\chi_{\rm up}$ is the largest $|\ch|$ allowed by the profile-likelihood confidence region. We fix each candidate splitting, refit the other parameters, and retain values whose log likelihood lies within $t_\chi/2$ of the best split fit. The second condition states that zero splitting remains allowed; the third requires every allowed splitting to be small. We call outcomes satisfying all three conditions EP-compatible. The allowance $b_\chi$ accounts for model and numerical shifts, and $\chi_*$ sets the target resolution, corresponding to mode separation $2\gamma_c\sqrt{\chi_*}$. Noise simulations assess the thresholds for the specified response family and observing window; definitions, profiles, and conditional calibrations are given in SM, Secs.~VII and IX.

\section{Waveform distance and resolution}
\label{sec:resolution}

To determine what makes splitting measurable, fix a signal with known nonzero splitting and find the EP waveform that best reproduces it. The remaining squared difference is
\begin{equation}
    d_{\EP}^2=\min_{\theta_{\EP}}
    \norm{h_{\rm split}-h_{\EP}(\theta_{\EP})}^2.
    \label{eq:distance}
\end{equation}
Here $h$ denotes the waveform in the chosen readout, and the norm weights samples and channels by their noise covariance $N$: $\norm{v}^2=v^\dagger N^{-1}v$. Only the candidate EP parameters $\theta_{\EP}$ vary: its center, excitation, and included ordinary responses. At fixed signal normalization, time window, and noise weighting, a larger $d_{\EP}^2$ makes splitting easier to distinguish. This noise-free distance measures the information remaining after refitting; noisy-data inference uses Eq.~\eqref{eq:criterion}.

Let $\epsilon=\sqrt{|\ch|}=|s_+-s_-|/(2\gamma_c)$ measure the mode separation. For a single response with nonzero $C$, the relative Jordan excitation $J/(\gamma_c C)$ compares the Jordan and constant terms over one decay time, independently of the overall amplitude. For the EP cluster alone with a freely fitted center, asymptotic matching gives $d_{\EP}^2\propto\epsilon^4$ when this ratio approaches a finite nonzero limit. Consider instead two split modes whose amplitudes approach finite, nonzero, unequal values with a nonzero sum. The relative Jordan excitation then decreases linearly with $\epsilon$. A center shift absorbs the leading quadratic time dependence in Eq.~\eqref{eq:response}, leaving an $O(\epsilon^3)$ residual and hence $d_{\EP}^2\propto\epsilon^6$. The derivation and other excitation paths are given in SM, Secs.~II and III.

The lower power leaves a larger difference at sufficiently small separation. At fixed discrimination strength, specified by a target value of the noise-free splitting evidence $T_\chi$, fourth- and sixth-order squared distances require signal-to-noise ratios (SNRs) $\rho\propto\epsilon^{-2}$ and $\rho\propto\epsilon^{-3}$: halving the separation costs four or eight times the SNR. Along the path with linearly decreasing relative Jordan excitation, the Jordan term can therefore be detected while the splitting remains unresolved. Its interpretation then depends on whether other allowed response components can absorb or reproduce the small distinguishing residual.

\section{Numerical setup and reference waveforms}
\label{sec:waveforms}

\begin{figure*}[t]
    \centering
    \subfloat[]{\includegraphics[width=.323\textwidth]{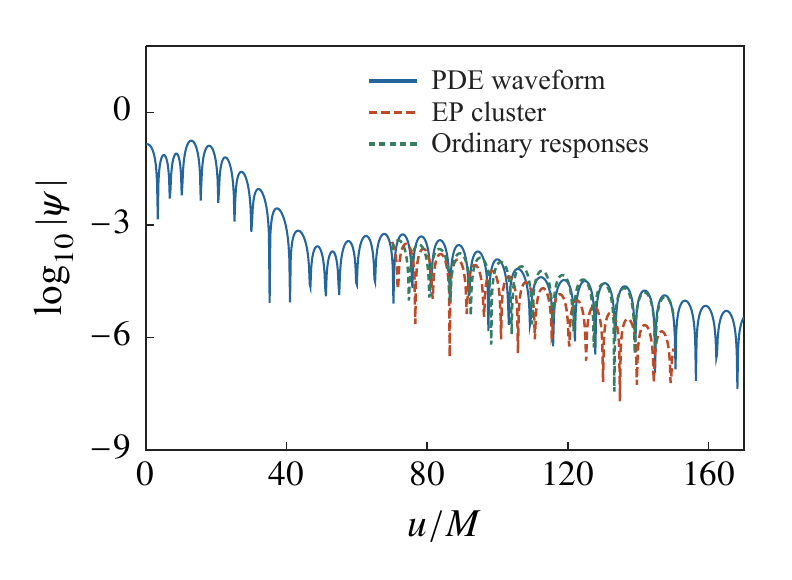}}
    \hfill
    \subfloat[]{\includegraphics[width=.323\textwidth]{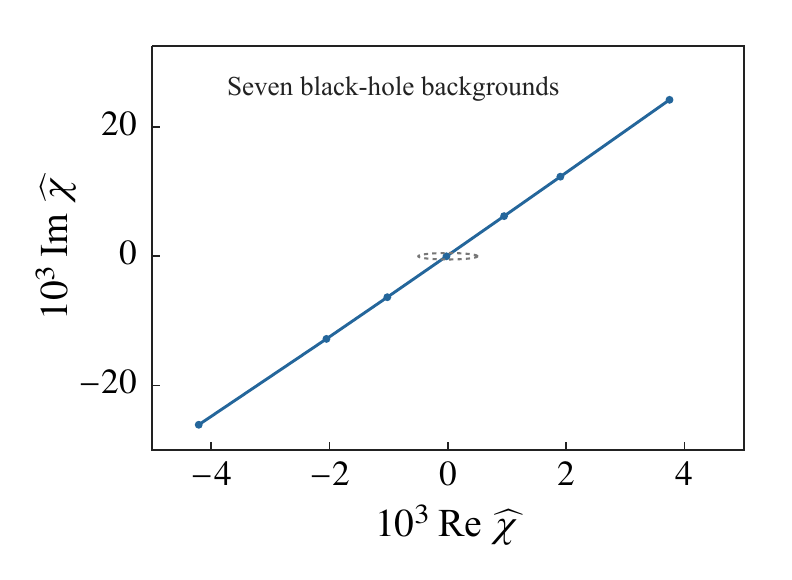}}
    \hfill
    \subfloat[]{\includegraphics[width=.323\textwidth]{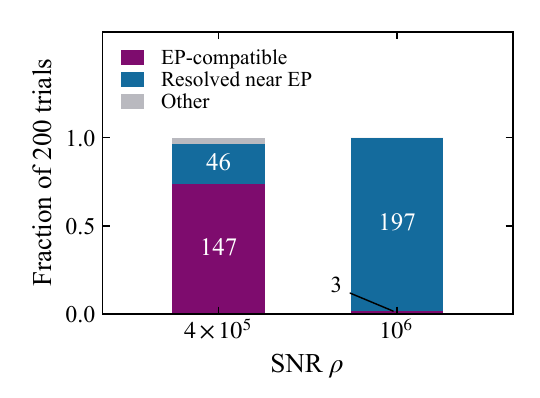}}
    \caption{Complete EMS waveforms and finite-resolution inference. (a) One of two perturbations with unit total energy at the nominal EP, observed at $r_*/M=60$ on $70\le u/M\le150$, where $u$ is retarded time. (b) Recovered splitting at $10^4(q-q_{\EP})=-4,-2,-1,0,1,2,4$, from top to bottom, using $80\le u/M\le160$; the circle marks the reference allowance $b_\chi=5\times10^{-4}$. (c) Two-source configuration with true $|\ch|\simeq1.02\times10^{-3}$: 200 Gaussian-noise trials per SNR, with two ordinary response terms, four tail terms, and the window of (a) (SM, Sec.~VI). EP-compatible denotes the full criterion. Resolved near EP denotes trials with $T_J>t_J$, $T_\chi>t_\chi$, and the same splitting upper-bound condition; gray denotes other outcomes. Thresholds are $t_J=15$, $t_\chi=10$, and $\chi_*=0.005$.}
    \label{fig:physical}
\end{figure*}

We generate complete waveforms by evolving a neutral scalar perturbation on an Einstein--Maxwell--scalar (EMS) BH \cite{Cheng2026}. Two-source fits combine waveforms generated by separately evolving two different finite-energy initial perturbations on the same background. The common evolution equation fixes their spectrum, while their excitation coefficients can differ. The reference EP, obtained numerically, has scalar--Maxwell coupling $\alpha\simeq0.79041$ and charge-to-mass ratio $q_{\EP}\equiv(Q/M)_{\EP}\simeq1.04219$. Figure~\ref{fig:physical}(a) displays a waveform and its fitted components; Fig.~\ref{fig:physical}(b) shows the splitting recovered on nearby backgrounds. The source construction and independent spectrum are given in SM, Secs.~VI and XV.

The reference fit contains the EP cluster, two ordinary response terms, and four tail terms on a finite retarded-time window. The ordinary terms describe damped responses outside the EP cluster; the faster term is an empirical finite-window component, without an independent QNM identification. Joint fits impose common frequencies and decay rates, with independent complex amplitudes for each source. We add correlated Gaussian noise and vary its level to set $\rho^2=\norm{h}^2$, using the same covariance in the likelihood. The window and thresholds are specified in Fig.~\ref{fig:physical}; fitting details are in SM, Sec.~VII. Comparisons of ordinary response models keep the signal and noise realizations fixed. Center-information comparisons keep the target waveforms and normalization fixed while varying the center constraints.

The complete scalar waveforms provide controlled near-EP signals for testing ordinary responses and fit freedom. An effective model of axial gravitational perturbations demonstrates the EP cluster response in tensor radiation \cite{YangBertiFranchini2025,MartelPoisson2005}. Simulated strain tests whether split and EP waveforms remain distinguishable after polarization readout and detector noise (SM, Secs.~XI and XII).

Finite resolution permits a neighborhood of the mathematical EP: at fixed $\alpha$, $|\ch|<0.005$ corresponds locally to $|q-q_{\EP}|\lesssim7.9\times10^{-5}$. Whether data identify coalescence also depends on their precision. In Fig.~\ref{fig:physical}(c), the true splitting is $|\ch|\simeq1.02\times10^{-3}$. At $\rho=4\times10^5$, 147 of 200 realizations satisfy Eq.~\eqref{eq:criterion}; at $\rho=10^6$, only 3 do, while 197 resolve the nonzero splitting within the target bound. The same near-EP spectrum is therefore compatible with coalescence at lower precision and distinguishable at higher precision. The parameter map and recovery errors are given in SM, Sec.~VI.

\section{Ordinary responses and model completeness}
\label{sec:ordinary}

\begin{figure*}[t]
    \centering
    \subfloat[]{\includegraphics[width=.323\textwidth]{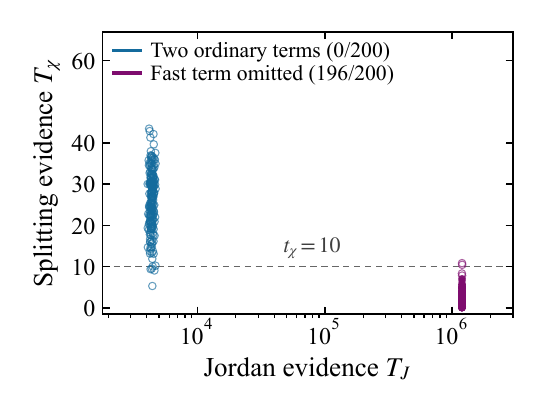}}
    \hfill
    \subfloat[]{\includegraphics[width=.323\textwidth]{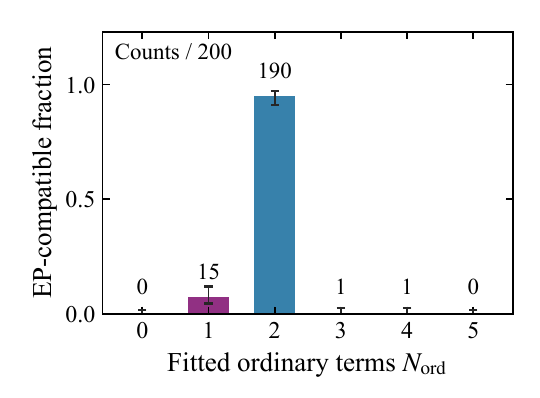}}
    \hfill
    \subfloat[]{\includegraphics[width=.323\textwidth]{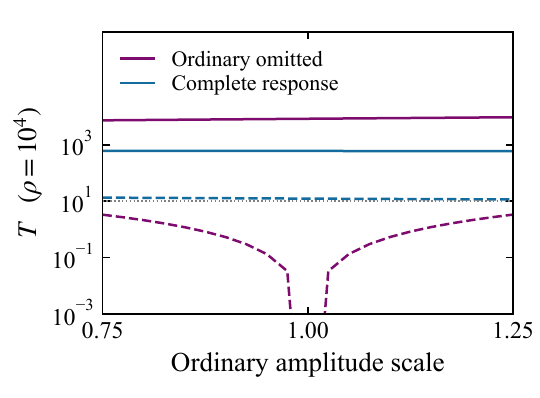}}
    \caption{Ordinary responses change EP inference. (a) Single-source split EMS signal at $\rho=2.3\times10^5$, fitted in 200 paired noise trials with the reference ranges of ordinary frequencies and decay rates (SM, Sec.~VII). All points satisfy $T_J>15$; the dashed line $T_\chi=10$ separates rejection of zero splitting above from compatibility below. Filled points are EP-compatible; parentheses give their counts. (b) EP-compatible fraction for two sources on the nominal EP at $\rho=4\times10^5$, with 200 trials per response count and 95\% Wilson intervals. Both panels use the thresholds in Fig.~\ref{fig:physical}. (c) Simulated tensor strain with true $\ch=-0.0256$ at $\rho=10^4$: in the incomplete fit, Jordan evidence grows with ordinary-component amplitude over the plotted range, while splitting evidence is minimized near the selected baseline amplitude. Solid and dashed curves show $T_J$ and $T_\chi$ (SM, Sec.~XI).}
    \label{fig:omission}
\end{figure*}

Omission can make a truly split signal appear to be an EP. On the split EMS background $q-q_{\EP}=2\times10^{-4}$, we select finite-energy initial data before adding noise, then fit the same waveform with two ordinary components or only the slower one. At $\rho=2.3\times10^5$, the reference fit rejects zero splitting in $196/200$ trials and gives no EP-compatible trials. Omitting the faster component instead gives EP-compatible results in $196/200$ paired trials [Fig.~\ref{fig:omission}(a)]. Restoring the response recovers splitting evidence within the specified ranges of ordinary frequencies and decay rates. Expanding those ranges lowers the noise-free splitting evidence $T_\chi$ below threshold; source, residual, and range controls are in SM, Sec.~IX.

The mechanism is coherent interference: the relative phase of an ordinary component can reinforce the Jordan-like part of the signal while canceling time dependence that distinguishes the split pair. Excluding this component attributes both effects to the EP cluster, increasing Jordan evidence and narrowing the splitting region around a biased estimate. The corresponding noise-free contrast persists over a region of nonzero area in parameters describing the unit-energy initial perturbation (SM, Sec.~IX). Coherent omission also biases tensor strain [Fig.~\ref{fig:omission}(c)].

Allowing more responses also costs resolution. Figure~\ref{fig:omission}(b) turns to the nominal EP background to test whether additional fit freedom still permits a tight splitting bound. We vary $N_{\rm ord}$, the number of fitted ordinary response terms, from zero to five; the EP cluster and tails are excluded from this count. At $\rho=4\times10^5$, two terms yield $190/200$ EP-compatible trials; three, four, and five yield $1/200$, $1/200$, and $0/200$. Additional unknown frequencies and amplitudes allow larger splittings to fit the data, widening the allowed region even when the best fit improves very little. A good fit with two terms therefore need not establish a splitting bound that survives other physically plausible response models (SM, Sec.~IX).

The required response content depends on the signal, window, and desired resolution. Comparing all eight retained subsets in strain containing three known ordinary components shows how omission bias depends on component identity, amplitude, and phase; the full noise-free model recovers the true splitting (SM, Sec.~IX). A local analytic overlap relation relates omission bias to lost resolution: an ordinary response overlapping with the waveform change caused by varying the splitting shifts the inferred splitting when omitted, and absorbs splitting information when fitted (SM, Sec.~X). Reliable inference must therefore account for this freedom and identify which additional measurements constrain it.

\section{Resolution with constrained centers}
\label{sec:center}

\begin{figure*}[t]
    \centering
    \subfloat[]{\includegraphics[width=.323\textwidth]{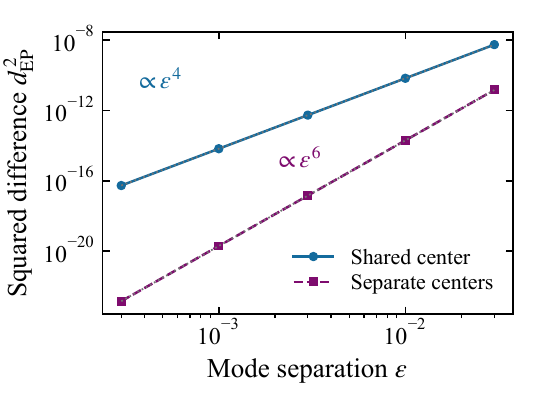}}
    \hfill
    \subfloat[]{\includegraphics[width=.323\textwidth]{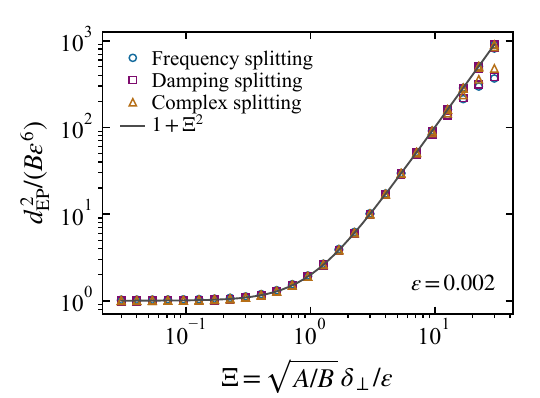}}
    \hfill
    \subfloat[]{\includegraphics[width=.323\textwidth]{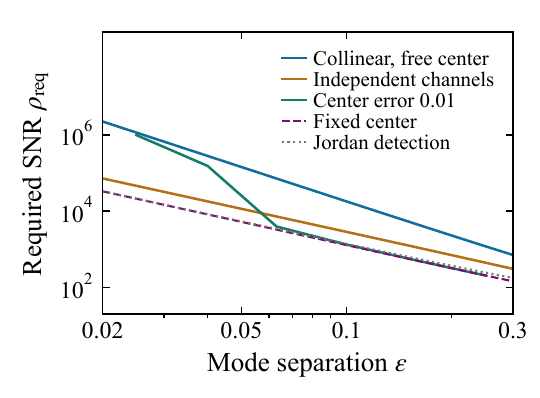}}
    \caption{Information that improves splitting resolution. (a) The same two targets, $c=(1,0)$, $J/\gamma_c=\epsilon(0.5,0.4)$, and $\ch=-\epsilon^2$, fitted with shared or separate EP centers; dotted lines show the asymptotic laws. (b) Squared distances normalized by $B\epsilon^6$ follow $1+\Xi^2$, where $\Xi=\sqrt{A/B}\,\delta_\perp/\epsilon$. (c) Required SNR at noise-free $T_\chi=10$: $f_c=150$ Hz, $Q=3$, and a polarization-resolving network with orthogonal antenna rows. Free-center fits have $T_\chi=\rho^2d_{\EP}^2/E_h$, with squared signal norm $E_h$. The center-uncertainty curve includes the external Gaussian penalty (SM, Sec.~XI), with standard deviation $0.01\gamma_c$ in angular frequency and decay rate. Collinear means proportional excitation; independently excited radiative channels have $\delta_\perp=0.4$. Jordan detection compares the single-exponential and EP models at threshold 15.}
    \label{fig:resolution}
\end{figure*}

Center adjustment removes the leading splitting dependence when relative Jordan excitation decreases linearly. Constraints preventing this compensation can preserve an $O(\epsilon^2)$ waveform difference and a fourth-order squared distance (SM, Secs.~III and IV). The constraint may come from another measurement or from consistency between distinct responses; its strength determines the magnitude and range of the improvement.

An independently known mean frequency and decay rate directly restrict the compensating shift. An exactly fixed center preserves the leading splitting term; finite measurement errors permit some adjustment. The useful precision is set by the size and direction of the required shift, which may involve frequency, decay rate, or both. The compensating shift becomes smaller as the mode separation decreases. Fixed finite precision improves resolution over a range of separations. At fixed discrimination strength, maintaining $\rho\propto\epsilon^{-2}$ toward coalescence requires improving that precision. Other modes or earlier radiation can constrain the center through a common remnant model, with correlations included in the joint likelihood. Controls with two, three, and five ordinary response terms retain substantial gains from finite center information (SM, Sec.~IV).

Different excitations of the same EP cluster share its frequencies and decay rates. Separate fits can cancel each response's leading mismatch by shifting its EP center. A joint fit must use the same center for every response. An overall complex amplitude change rescales both components and leaves the compensating center unchanged. Write this center as $s_c+\gamma_c\epsilon a$, where $a$ is complex, and set $c_P=C_P$, $k_P=J_P/(\gamma_c\epsilon)$, and $\ch=\epsilon^2\zeta$ with $|\zeta|=1$. For complex responses in white noise on $x=\gamma_c\tau\ge0$, with squared norm $\sum_P\int_0^\infty |H_P(x)|^2\,dx$, we derive
\begin{equation}
    d_{\EP}^2=\frac{\epsilon^4}{32}\min_{a\in\mathbb C}
    \sum_P\left|(a^2+\zeta)c_P-2ak_P\right|^2+o(\epsilon^4).
    \label{eq:shared_center}
\end{equation}
Here $c$ and $k$ collect the coefficients across responses, and each expression inside the modulus controls the corresponding leading mismatch after refitting.
All leading mismatches vanish only if $k$ and $c$ are proportional. Different relative compositions of the Jordan and constant terms therefore preserve a fourth-order term. This use of excitation diversity is related to resolution gains in multiple-snapshot spectral estimation \cite{LiMultiSnapshot2021,LiuMultiSnapshot2022}. Figure~\ref{fig:resolution}(a) compares shared and separate centers for identical targets and normalization; the latter recover sixth order. SM, Secs.~III and VIII give the proof, weighted extension, and complete EMS realization.

Within one event, the source must supply these composition differences and the readout must retain them. Multiple fixed readouts, such as detector responses or polarization projections, form linear combinations of the same excited EP cluster. Along the path with linearly decreasing relative Jordan excitation, their leading responses admit a common compensating center and generically retain sixth order. Independently excited axial components with the same $\ell$ and different $m$ on a spherical background obey the same radial equation. They share a spectrum but can have different relative excitation compositions. These differences can appear in distinguishable polarizations at suitable viewing directions and preserve the fourth-order term if the detector network retains them. The radiative construction and proofs are in SM, Secs.~IV and XII.

The excitation contrast sets how much information survives. Let $\delta_\perp$ measure the part of $k$ not proportional to $c$, normalized by $\norm{c}$. It measures differences in the relative mixtures of the constant and Jordan contributions across responses. For nearly proportional compositions in the complex white-noise benchmark, we derive the generic uniform result
\begin{equation}
    d_{\EP}^2\simeq A\,\delta_\perp^2\epsilon^4+B\,\epsilon^6,
    \qquad A,B>0.
    \label{eq:uniform}
\end{equation}
The coefficients depend on the excitation path and normalization. Figure~\ref{fig:resolution}(b) plots the squared-distance gain $d_{\EP}^2/(B\epsilon^6)$ relative to the sixth-order baseline. With $\Xi=\sqrt{A/B}\,\delta_\perp/\epsilon$, the ratio of fourth- to sixth-order contributions is $\Xi^2$. The improvement is small for $\Xi\ll1$, the terms are comparable near $\Xi=1$, and the fourth-order term dominates for $\Xi\gg1$. At fixed signal normalization and discrimination strength, the corresponding required SNR scales as $\rho\propto\epsilon^{-3}$ and $\rho\propto\epsilon^{-2}\delta_\perp^{-1}$ in the two limiting regimes. SM, Secs.~III and X give the uniform remainder and noise-weighted extensions to finite windows, real strain, and ordinary-response fitting.

Figure~\ref{fig:resolution}(c) compares the observational cost in Advanced LIGO design noise \cite{LIGODesignT1800044}. At $f_c=\Omega_c/(2\pi)=150$ Hz, $Q=\Omega_c/(2\gamma_c)=3$, and $\epsilon=0.16$, a noise-free $T_\chi=10$ requires $\rho\simeq4.4\times10^3$ for proportional responses, $1.1\times10^3$ for distinct radiative excitations, or $5.0\times10^2$ with the specified center measurement. The last value uses a joint likelihood including the external Gaussian measurement. These benchmarks quantify different gains from restricting the same compensation; the improvement depends on the information retained after detector readout and response fitting. Detection fractions from noisy realizations, including randomized center measurements, are given in SM, Sec.~XI.

\section{Discussion and conclusions}
\label{sec:conclusion}

Identifying mode coalescence in finite ringdown data requires a splitting bound that remains informative when the ordinary responses are modeled. Coherent ordinary responses can strengthen Jordan evidence and produce a spuriously tight bound when omitted; allowing their parameters to vary controls that bias but can weaken resolution. The analytic overlap relation explains both effects: a response aligned with the splitting-dependent waveform can bias its estimate or reduce its distinguishability. For linearly decreasing relative Jordan excitation, a free center further suppresses the squared waveform difference to sixth order in the normalized mode separation $\epsilon$. Constraints that prevent this compensation restore a contribution proportional to $\epsilon^4$, with its coefficient and useful range set by excitation differences, measurement precision, and radiative readout. Thus independent information can improve sensitivity while retaining the response freedom needed for reliable inference.

For a specific merger, the remnant spectrum relates other measurements to the target center, while the excitation amplitudes and their correlations determine which independent radiation components are available. The viewing direction and detector network select the differences that survive in the strain. Combining these physical inputs with the derived resolution laws and response-model comparisons determines how precisely an event can distinguish coalesced modes from a nearby split pair, and thereby probe the degrees of freedom governing BH relaxation.

\section*{Acknowledgements}
\label{sec:acknowledgements}
Z.-X. Z is supported by the Pilot Scheme of Talent Training in Basic Sciences (Boling Class of Physics, Nankai University), Ministry of Education.

\nocite{Badeau2008,Cavalcante2024,Cheng2026,CutlerVallisneri2007,DeAmicis2026,Drton2009,GundlachPricePullin1994,IsiFarr2021,Kato1966,LIGODesignT1800044,Leaver1985,LiMultiSnapshot2021,LiuMultiSnapshot2022,MaCleaning2023,MaFilters2022,Macedo2025,MartelPoisson2005,MorikawaOgawaHirose2026,ScharfFriedlander1994,YangBertiFranchini2025}

\bibliography{references}

@misc{SupplementalMaterial,
 title={Supplemental Material for {Identifying Exceptional Points in Black-Hole Ringdowns}},
 year={2026},
 note={For response-distance derivations, radiative excitation and center information, finite-energy black-hole waveforms, ordinary-response interference, detector likelihoods, and additional spectral results. Data and numerical programs accompany the Supplemental Material}}

@article{BertiCardosoStarinets2009,
 author={Berti, Emanuele and Cardoso, Vitor and Starinets, Andrei O.},
 title={Quasinormal modes of black holes and black branes},
 journal={Classical and Quantum Gravity}, volume={26}, pages={163001}, year={2009}, doi={10.1088/0264-9381/26/16/163001},
 eprint={0905.2975}, archivePrefix={arXiv}, primaryClass={gr-qc}, url={https://arxiv.org/abs/0905.2975}}

@article{Isi2019,
 author={Isi, Maximiliano and Giesler, Matthew and Farr, Will M. and Scheel, Mark A. and Teukolsky, Saul A.},
 title={Testing the no-hair theorem with {GW150914}},
 journal={Physical Review Letters}, volume={123}, pages={111102}, year={2019}, doi={10.1103/PhysRevLett.123.111102},
 eprint={1905.00869}, archivePrefix={arXiv}, primaryClass={gr-qc}, url={https://arxiv.org/abs/1905.00869}}

@article{LVK2026GW250114,
 author={Abac, A. G. and others}, collaboration={LIGO Scientific, Virgo, and KAGRA},
 title={Black hole spectroscopy and tests of general relativity with {GW250114}},
 journal={Physical Review Letters}, volume={136}, pages={041403}, year={2026}, doi={10.1103/6c61-fm1n},
 eprint={2509.08099}, archivePrefix={arXiv}, primaryClass={gr-qc}, url={https://arxiv.org/abs/2509.08099}}

@book{Kato1966,
 author={Kato, Tosio}, title={Perturbation Theory for Linear Operators},
 publisher={Springer-Verlag}, address={Berlin}, series={Grundlehren der mathematischen Wissenschaften}, volume={132}, year={1966}, doi={10.1007/978-3-662-12678-3}}

@article{MiriAlu2019,
 author={Miri, Mohammad-Ali and Al{\`u}, Andrea}, title={Exceptional points in optics and photonics},
 journal={Science}, volume={363}, pages={eaar7709}, year={2019}, doi={10.1126/science.aar7709}}

@article{Cavalcante2024,
 author={Cavalcante, Jo{\~a}o Paulo and Richartz, Maur{\'i}cio and Carneiro da Cunha, Bruno},
 title={Exceptional point and hysteresis in perturbations of {Kerr} black holes},
 journal={Physical Review Letters}, volume={133}, pages={261401}, year={2024}, doi={10.1103/PhysRevLett.133.261401},
 eprint={2407.20850}, archivePrefix={arXiv}, primaryClass={gr-qc}, url={https://arxiv.org/abs/2407.20850}}

@article{YangBertiFranchini2025,
 author={Yang, Yiqiu and Berti, Emanuele and Franchini, Nicola},
 title={Black hole quasinormal mode resonances}, journal={Physical Review Letters},
 volume={135}, pages={201401}, year={2025}, doi={10.1103/hfv8-n444}, eprint={2504.06072}, archivePrefix={arXiv}, primaryClass={gr-qc}, url={https://arxiv.org/abs/2504.06072}}

@misc{Cheng2026,
 author={Cheng, Lang and Guo, Xiaobo and Li, Yuhan and Tao, Jun and Wang, Peng},
 title={Exceptional Points in Quasinormal Spectra of Hairy Black Holes},
 year={2026}, eprint={2603.22261}, archivePrefix={arXiv}, primaryClass={gr-qc}, url={https://arxiv.org/abs/2603.22261}}

@article{Macedo2025,
 author={Macedo, Rodrigo P. and Katagiri, Takuya and Kubota, Kei-ichiro and Motohashi, Hayato},
 title={Exceptional points and resonance in black hole ringdown},
 journal={Physical Review D}, volume={113}, pages={L121504}, year={2026}, doi={10.1103/5w5d-kr25},
 eprint={2512.02110}, archivePrefix={arXiv}, primaryClass={gr-qc}, url={https://arxiv.org/abs/2512.02110}}

@misc{MorikawaOgawaHirose2026,
 author={Morikawa, Okuto and Ogawa, Shoya and Hirose, Takuya},
 title={{Riesz--Laurent} representation of black-hole scattering and sourced response at exceptional points},
 year={2026}, eprint={2608.14752}, archivePrefix={arXiv}, primaryClass={gr-qc}, url={https://arxiv.org/abs/2608.14752}}

@article{ImafukuOshitaTakeda2026,
 author={Imafuku, Hayato and Oshita, Naritaka and Takeda, Hiroki},
 title={Detectability of avoided crossings in black hole ringdowns},
 journal={Physical Review D}, volume={114}, pages={064014},
 year={2026}, doi={10.1103/lpm5-28jm}, eprint={2605.16199}, archivePrefix={arXiv}, primaryClass={gr-qc}, url={https://arxiv.org/abs/2605.16199}}

@article{Fuchs2014,
 author={Fuchs, J. and Main, J. and Cartarius, H. and Wunner, G.},
 title={Harmonic inversion analysis of exceptional points in resonance spectra},
 journal={Journal of Physics A: Mathematical and Theoretical}, volume={47}, pages={125304}, year={2014}, doi={10.1088/1751-8113/47/12/125304},
 eprint={1402.4032}, archivePrefix={arXiv}, primaryClass={nlin.CD}, url={https://arxiv.org/abs/1402.4032}}

@article{DyerMoore2026,
 author={Dyer, Richard and Moore, Christopher J.}, title={Quasinormal mode content of binary black hole ringdowns},
 journal={Physical Review Letters}, volume={136}, pages={191403}, year={2026}, doi={10.1103/ptmd-rz1t},
 eprint={2510.13954}, archivePrefix={arXiv}, primaryClass={gr-qc}, url={https://arxiv.org/abs/2510.13954}}

@article{DeAmicis2026,
 author={De Amicis, Marina and Cannizzaro, Enrico and Carullo, Gregorio and Sberna, Laura},
 title={Dynamical quasinormal mode excitation}, journal={Physical Review D}, volume={113}, pages={024048}, year={2026}, doi={10.1103/bgjv-lvyd},
 eprint={2506.21668}, archivePrefix={arXiv}, primaryClass={gr-qc}, url={https://arxiv.org/abs/2506.21668}}

@article{Leaver1985,
 author={Leaver, Edward W.}, title={An analytic representation for the quasi-normal modes of {Kerr} black holes},
 journal={Proceedings of the Royal Society of London A}, volume={402}, pages={285--298}, year={1985}, doi={10.1098/rspa.1985.0119}}

@article{GundlachPricePullin1994,
 author={Gundlach, Carsten and Price, Richard H. and Pullin, Jorge},
 title={Late-time behavior of stellar collapse and explosions. {I}. Linearized perturbations},
 journal={Physical Review D}, volume={49}, pages={883--889}, year={1994}, doi={10.1103/PhysRevD.49.883},
 eprint={gr-qc/9307009}, archivePrefix={arXiv}, url={https://arxiv.org/abs/gr-qc/9307009}}

@article{LiuMultiSnapshot2022,
 author={Liu, Ping and Yu, Sanghyeon and Sabet, Ola and Pelkmans, Lucas and Ammari, Habib},
 title={Mathematical Foundation of Sparsity-based Multi-snapshot Spectral Estimation},
 journal={Applied and Computational Harmonic Analysis}, volume={73}, pages={101673},
 year={2024}, doi={10.1016/j.acha.2024.101673}, eprint={2202.11189}, archivePrefix={arXiv}, url={https://arxiv.org/abs/2202.11189}}

@article{LiMultiSnapshot2021,
 author={Li, Weilin and Zhu, Zengying and Gao, Weiguo and Liao, Wenjing},
 title={Stability and Super-resolution of {MUSIC} and {ESPRIT} for Multi-snapshot Spectral Estimation},
 journal={IEEE Transactions on Signal Processing}, volume={70}, pages={4555--4570},
 year={2022}, doi={10.1109/TSP.2022.3204454}, eprint={2105.14304}, archivePrefix={arXiv}, url={https://arxiv.org/abs/2105.14304}}

@article{CutlerVallisneri2007,
 author={Cutler, Curt and Vallisneri, Michele},
 title={{LISA} detections of massive black hole inspirals: Parameter extraction errors due to inaccurate template waveforms},
 journal={Physical Review D}, volume={76}, pages={104018}, year={2007},
 doi={10.1103/PhysRevD.76.104018}, eprint={0707.2982}, archivePrefix={arXiv}, primaryClass={gr-qc}, url={https://arxiv.org/abs/0707.2982}}

@article{Drton2009,
 author={Drton, Mathias}, title={Likelihood ratio tests and singularities},
 journal={The Annals of Statistics}, volume={37}, pages={979--1012}, year={2009}, doi={10.1214/07-AOS571},
 eprint={math/0703360}, archivePrefix={arXiv}, url={https://arxiv.org/abs/math/0703360}}

@article{BidSchomerus2024,
 author={Bid, Subhajyoti and Schomerus, Henning},
 title={Uniform response theory of {non-Hermitian} systems: {Non-Hermitian} physics beyond the exceptional point},
 journal={Physical Review Research}, volume={7}, pages={023062}, year={2025},
 doi={10.1103/PhysRevResearch.7.023062}, eprint={2412.11932}, archivePrefix={arXiv}, primaryClass={quant-ph}, url={https://arxiv.org/abs/2412.11932}}

@article{Badeau2008,
 author={Badeau, Roland and David, Bertrand and Richard, Ga{\"e}l},
 title={{Cram{\'e}r--Rao} bounds for multiple poles and coefficients of quasi-polynomials in colored noise},
 journal={IEEE Transactions on Signal Processing}, volume={56}, number={8},
 pages={3458--3467}, year={2008}, doi={10.1109/TSP.2008.921719}}

@article{ScharfFriedlander1994,
 author={Scharf, Louis L. and Friedlander, Benjamin},
 title={Matched subspace detectors}, journal={IEEE Transactions on Signal Processing},
 volume={42}, number={8}, pages={2146--2157}, year={1994}, doi={10.1109/78.301849}}

@article{MaFilters2022,
 author={Ma, Sizheng and Mitman, Keefe and Sun, Ling and Deppe, Nils and H{\'e}bert, Fran{\c c}ois and Kidder, Lawrence E. and Moxon, Jordan and Throwe, William and Vu, Nils L. and Chen, Yanbei},
 title={Quasinormal-mode filters: A new approach to analyze the gravitational-wave ringdown of binary black-hole mergers},
 journal={Physical Review D}, volume={106}, pages={084036}, year={2022},
 doi={10.1103/PhysRevD.106.084036}, eprint={2207.10870}, archivePrefix={arXiv}, primaryClass={gr-qc}, url={https://arxiv.org/abs/2207.10870}}

@article{MaCleaning2023,
 author={Ma, Sizheng and Sun, Ling and Chen, Yanbei},
 title={Black hole spectroscopy by mode cleaning},
 journal={Physical Review Letters}, volume={130}, pages={141401}, year={2023},
 doi={10.1103/PhysRevLett.130.141401}, eprint={2301.06705}, archivePrefix={arXiv}, primaryClass={gr-qc}, url={https://arxiv.org/abs/2301.06705}}

@misc{IsiFarr2021,
 author={Isi, Maximiliano and Farr, Will M.},
 title={Analyzing black-hole ringdowns},
 year={2021}, eprint={2107.05609}, archivePrefix={arXiv}, primaryClass={gr-qc}, url={https://arxiv.org/abs/2107.05609}}

@article{MartelPoisson2005,
 author={Martel, Karl and Poisson, Eric},
 title={Gravitational perturbations of the {Schwarzschild} spacetime: A practical covariant and gauge-invariant formalism},
 journal={Physical Review D}, volume={71}, pages={104003}, year={2005}, doi={10.1103/PhysRevD.71.104003}, eprint={gr-qc/0502028}, archivePrefix={arXiv}}

@misc{LIGODesignT1800044,
 author={Barsotti, Lisa and Gras, Slawomir and Evans, Matthew and Fritschel, Peter},
 title={The updated {Advanced LIGO} design curve},
 howpublished={LIGO Technical Note T1800044-v5},
 year={2018},
 url={https://dcc.ligo.org/LIGO-T1800044-v5/public},
 note={{ASD} data file aLIGODesign.txt; accessed 2026-09-13}}

@article{Motohashi2025,
 author={Motohashi, Hayato},
 title={Resonant Excitation of Quasinormal Modes of Black Holes},
 journal={Physical Review Letters}, volume={134}, number={14}, pages={141401}, year={2025},
 doi={10.1103/PhysRevLett.134.141401}, eprint={2407.15191}, archivePrefix={arXiv}, primaryClass={gr-qc}, url={https://arxiv.org/abs/2407.15191}}

@article{TakahashiMotohashiTakahashi2025,
 author={Takahashi, Takuya and Motohashi, Hayato and Takahashi, Kazufumi},
 title={Resonance of black hole quasinormal modes in coupled systems},
 journal={Physical Review D}, volume={112}, pages={064006},
 year={2025}, doi={10.1103/59tb-2x9r}, eprint={2505.03883}, archivePrefix={arXiv}, primaryClass={gr-qc},
 url={https://arxiv.org/abs/2505.03883}}
\clearpage
\setcounter{dbltopnumber}{2}
\onecolumngrid
\begin{center}
{\large\bfseries Supplemental Material}\\[4pt]
Identifying Exceptional Points in Black Hole Ringdowns\\[4pt]
Zhen-Xiao Zhang, Chen Lan, and Yi Wang
\end{center}
\twocolumngrid
\setcounter{section}{0}
\setcounter{equation}{0}
\setcounter{figure}{0}
\setcounter{table}{0}
\renewcommand{\thesection}{\Roman{section}}
\renewcommand{\theequation}{S\arabic{equation}}
\renewcommand{\thefigure}{S\arabic{figure}}
\renewcommand{\thetable}{S\arabic{table}}
\renewcommand{\theHfigure}{S\arabic{figure}}
\renewcommand{\theHtable}{S\arabic{table}}
\renewcommand{\theHequation}{S\arabic{equation}}
\renewcommand{\theHsection}{S\arabic{section}}
\raggedbottom

\section{EP cluster response and physical excitation}
Let $\mathcal A$ generate free linear evolution, and let $P=(2\pi\ii)^{-1}\oint(z-\mathcal A)^{-1}\dd z$ project onto an isolated two-dimensional EP cluster \cite{Kato1966,MorikawaOgawaHirose2026}. The contour encloses only the EP cluster; this projector isolates its contribution to the evolution. On this subspace write $\mathcal A_c=s_c I+D$, $\operatorname{tr}D=0$. Cayley--Hamilton gives $D^2=\chi I$, $\chi=-\det D$, and hence
\begin{equation}
    \begin{aligned}
        e^{\mathcal A_c\tau}&=e^{s_c\tau}\big[\CC_\chi(\tau)I+\SSS_\chi(\tau)D\big],\\
        (z-\mathcal A_c)^{-1}&=\frac{(z-s_c)I+D}{(z-s_c)^2-\chi},
    \end{aligned}
    \label{eq:s_resolvent}
\end{equation}
The inverse operator $(z-\mathcal A_c)^{-1}$ is the resolvent, which describes the response in the complex-frequency domain. Here $\CC_\chi=\cosh(\sqrt\chi\,\tau)$ and $\SSS_\chi=\sinh(\sqrt\chi\,\tau)/\sqrt\chi$. For initial excitation $u_j$ and readout $v_j$, the coefficients are $C_j=v_jPu_j$ and $J_j=v_jDPu_j$. At an EP, $D^2=0$ with $D\ne0$, and the double-pole coefficient in the measured response is $J_j$. This gives the source dependence of the polynomial ringdown discussed in Refs.~\cite{Macedo2025,MorikawaOgawaHirose2026}.

The transfer function for one source and one measured response is
\begin{equation}
    H_c(z)=\frac{C(z-s_c)+J}{(z-s_c)^2-\chi},\qquad
    \mathcal R=J^2-C^2\chi.
    \label{eq:s_transfer}
\end{equation}
Its minimal order, the denominator degree after canceling common factors, is two when $\mathcal R\ne0$. A visible second-order EP has $\chi=0$ and $J\ne0$. The unforced EP cluster obeys $h_c''-2s_ch_c'+(s_c^2-\chi)h_c=0$. On any interval with $W=h_ch_c''-(h_c')^2\ne0$, the center and splitting follow from
\begin{equation}
    \begin{aligned}
        W&=-\mathcal R e^{2s_c\tau},\\
        s_c&=\frac{h_ch_c'''-h_c'h_c''}{2W},\qquad
        \chi=s_c^2+\frac{(h_c'')^2-h_c'h_c'''}{W}.
    \end{aligned}
\end{equation}
In sampled data these quantities are estimated by fitting the response, rather than differentiating the waveform.

A change of reference time by $\Delta$ transforms the amplitudes as
\begin{equation}
    \begin{pmatrix}C_\Delta\\J_\Delta\end{pmatrix}
        =e^{s_c\Delta}
        \begin{pmatrix}\CC_\chi(\Delta)&\SSS_\chi(\Delta)\\
            \chi\SSS_\chi(\Delta)&\CC_\chi(\Delta)\end{pmatrix}
            \begin{pmatrix}C\\J\end{pmatrix}.
                \label{eq:s_transport}
            \end{equation}
The splitting and minimal order are invariant, with $\mathcal R_\Delta=e^{2s_c\Delta}\mathcal R$. Excitation ratios are always specified at a common reference. The present waveforms evolve from initial data without further forcing. For a driven problem the source enters through the causal convolution; in particular, a resonant source acting on a simple pole can itself produce $\tau e^{s_c\tau}$. Dynamical excitation in BH relaxation is discussed in Ref.~\cite{DeAmicis2026}.

\section{Distance from weak to finite Jordan excitation}
We first use dimensionless time $x=\gamma\tau$, remove the common carrier, and define the complex white-noise norm for positive times $\norm{h}^2=\int_0^\infty\sum_j|h_j(x)|^2\dd x$. The distance $d_\EP$ is minimized over the complex EP center and the complex amplitudes fitted separately to each record, where a record is one measured response. For a single response with $c=C=1$, $j=J/\gamma=\kappa\ge0$, and $\ch=-\epsilon^2$,
\begin{equation}
    h=e^{-x}\left[\cos(\epsilon x)+\kappa\frac{\sin(\epsilon x)}\epsilon\right].
\end{equation}
An EP response matching the first three Taylor coefficients has the form $h_\EP=e^{(-1+\kappa-z)x}(1+zx)$ with $z^2=\kappa^2+\epsilon^2$. The third-derivative difference at $x=0$ is $2(\kappa^2+\epsilon^2)(\kappa-z)$. Subsequent adjustments of the center and amplitudes project the cubic mismatch off the lower time moments. The required residual polynomials are
\begin{equation}
    \begin{aligned}
        R_2&=x^2-2x+\tfrac12,&\norm{e^{-x}R_2}^2&=\tfrac18,\\
        R_3&=x^3-\tfrac92x^2+\tfrac92x-\tfrac34,
        &\norm{e^{-x}R_3}^2&=\tfrac9{32}.
    \end{aligned}
    \label{eq:s_polynomials}
\end{equation}
Choosing the closer branch $z=\sqrt{\kappa^2+\epsilon^2}$ gives
\begin{equation}
    \frac{d_\EP^2}{E_0}
    =\frac{\epsilon^4(\kappa^2+\epsilon^2)^2}
    {16\big(\sqrt{\kappa^2+\epsilon^2}+\kappa\big)^2}[1+o(1)],
    \qquad E_0=\tfrac12.
    \label{eq:s_uniform}
\end{equation}
This leading-order expression applies to bounded nonnegative $\kappa$ as $\epsilon\to0$. For a finite positive limiting $\kappa$, linearizing in the EP fit parameters gives $d_\EP^2/E_0\sim\kappa^2\epsilon^4/64$. For $\kappa=O(\epsilon)$, the quadratic-matching centers move by $O(\epsilon)$ and the surviving cubic residual gives
\begin{equation}
    \kappa=a\epsilon:\qquad
    \frac{d_\EP^2}{E_0}\sim
    \frac{(a^2+1)^2}{16\big(\sqrt{a^2+1}+a\big)^2}\epsilon^6.
\end{equation}
The intermediate regime $\epsilon\ll\kappa\ll1$ is where these expansions overlap. Higher time moments and tangent corrections are smaller than the displayed leading term in each regime. Thus $\kappa=a\epsilon^\beta$ gives exponent $\min(4+2\beta,6)$. At $\kappa=0$, the coefficient is $1/16$.

The reference energy $E_0$ in Eq.~\eqref{eq:s_uniform} is distinct from the actual signal norm,
\begin{equation}
    E_h=\norm{h}^2=\frac{\kappa^2+2\kappa+2+\epsilon^2}{4(1+\epsilon^2)}.
\end{equation}
The squared distance normalized by signal energy is $d_\EP^2/E_h$; multiplying it by $\rho^2$ gives the squared separation in noise units. This normalization changes finite coefficients while preserving the stated powers for the bounded paths.

\par\medskip\noindent
\begin{minipage}{\columnwidth}
 \centering
 \includegraphics[width=\columnwidth]{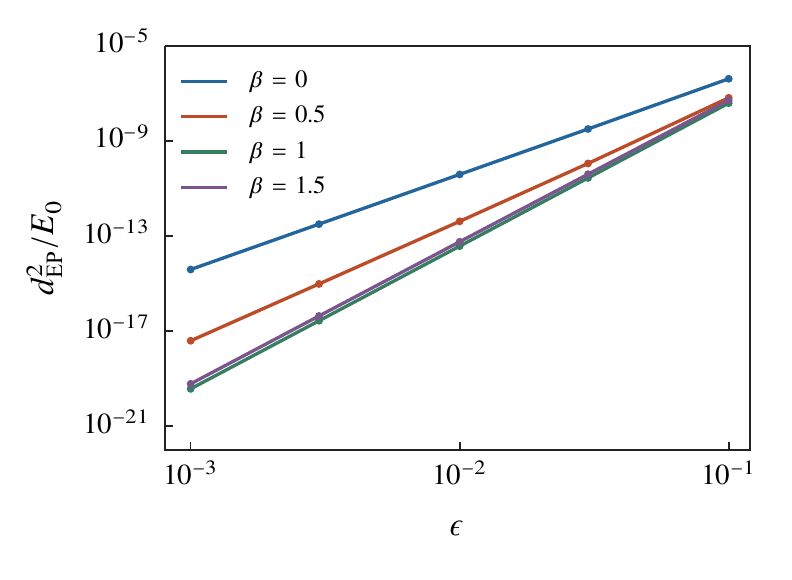}
 \captionof{figure}{Single-response distances for $\kappa=\tfrac12\epsilon^\beta$. Symbols are distances minimized over the full complex EP center and amplitudes. The asymptotic powers for $\beta=0,1/2,1,3/2$ are four, five, six, and six.}
 \label{fig:s_scalar}
\end{minipage}
\par\medskip

Finite-separation distances follow from the exponential inner product $\langle e^{s_i x},e^{s_k x}\rangle=(-\bar s_i-s_k)^{-1}$, known as the Cauchy Gram kernel. For $h_j=\sum_i A_{ji}e^{s_ix}$ and a fixed candidate center $q$, let $B_q(s)=[(s-q)/(s+\bar q)]^2$. Projection onto $e^{qx},xe^{qx}$ gives
\begin{equation}
    d^2(q)=\sum_{j,i,k}\frac{\bar A_{ji}A_{jk}\,
    \overline{B_q(s_i)}B_q(s_k)}{-\bar s_i-s_k}.
    \label{eq:s_cauchy}
\end{equation}
Minimizing over stable complex $q$ yields the curves in Fig.~\ref{fig:s_scalar} and the Letter. Distances at finite separation are evaluated at 65--75-digit precision, continuing the nearby quadratic-matching branches and retaining the minimum.

\section{The common-center condition and its crossover}
Let $c,k\in\mathbb C^m$, $c\ne0$, $j=\epsilon k$, and $\ch=\epsilon^2\zeta$ with $|\zeta|=1$. For an EP center shifted by $a\epsilon$, first-order matching fixes its linear coefficient to $\epsilon(k-ac)$. The second-derivative mismatch is then
\begin{equation}
    \epsilon^2 r(a),\qquad r(a)=(a^2+\zeta)c-2ak.
    \label{eq:s_obstruction}
\end{equation}
Independent record amplitudes remove the constant and linear moments. Using $R_2$ in Eq.~\eqref{eq:s_polynomials}, the remaining distance is
\begin{equation}
    d_\EP^2=\frac{\epsilon^4}{32}\min_{a\in\mathbb C}\norm{r(a)}^2+o(\epsilon^4).
    \label{eq:s_shared}
\end{equation}
Because $r(0)=\zeta c\ne0$, vanishing mismatch requires $a\ne0$ and $k=(a^2+\zeta)c/(2a)$. Conversely, when $k=\lambda c$, the scalar quadratic $a^2-2\lambda a+\zeta=0$ has a root. Complex collinearity is therefore the necessary and sufficient condition for the coefficient of the fourth-order squared distance to vanish. For noncollinear $c,k$, the residual grows as $|a|^2$ at large $|a|$ and has a strictly positive minimum.

Multiple-snapshot spectral-estimation analyses relate resolution to the information carried by different amplitude vectors \cite{LiMultiSnapshot2021,LiuMultiSnapshot2022}. Here the comparison is between a split response and its closest polynomial EP response, with their shared center profiled. Equation~\eqref{eq:s_shared} identifies the specific obstruction in this problem: distinct relative excitation compositions require incompatible leading center shifts. The resulting fourth- and sixth-order distances concern this asymptotic model comparison; the spectral-estimation bounds in those works address different estimators and signal classes.

The control with separately fitted centers uses the same two target waveforms as the $\beta=1$ curve in the Letter: $h_1=e^{-x}[\cos(\epsilon x)+\tfrac12\sin(\epsilon x)]$ and $h_2=\tfrac25e^{-x}\sin(\epsilon x)$. Its distance is $d_{\rm ind}^2=\sum_j\min_{q_j,A_j,B_j}\norm{h_j-e^{q_jx}(A_j+B_jx)}^2$, evaluated record by record with Eq.~\eqref{eq:s_cauchy}. The closest centers are $q_1=-1+(1-\sqrt5)\epsilon/2+O(\epsilon^2)$ and $q_2=-1-3\epsilon^2/4+O(\epsilon^4)$, giving $d_{\rm ind}^2\sim(1907-625\sqrt5)\epsilon^6/25600\simeq0.0199007\epsilon^6$. For a shared center the minimum occurs at real $a$, reducing to $\min_a[(a^2-a-1)^2+16a^2/25]/32=0.00671806$ at $a\simeq-0.541008$; hence $d_\EP^2\sim0.00671806\epsilon^4$. Both comparisons have the same total signal norm, tending to $1/2$, and the same white half-line inner product. Their different powers result entirely from the compatibility of the fitted centers.

For example, $\zeta=-1$, $c=(1,0)$, and $k=(0,K)$ with $K$ real give
\begin{equation}
    \min_a\norm{r(a)}^2=
    \begin{cases}4K^2(1-K^2),&K^2\le1/2,\\1,&K^2\ge1/2.\end{cases}
    \end{equation}
When $j=\epsilon^\beta k$ with $\beta>1$, define $k_\perp=k-c(c^\dagger k)/\norm{c}^2$. The channel-orthogonal quadratic residual and channel-parallel cubic residual give the combined leading-order expression
\begin{equation}
    d_\EP^2=\frac{\epsilon^{2\beta+2}}8\norm{k_\perp}^2
    +\frac{\epsilon^6}{32}\norm{c}^2
    +o(\epsilon^{2\beta+2}+\epsilon^6).
    \label{eq:s_vanishing}
\end{equation}
Thus the leading power is $2\beta+2$ for $1<\beta<2$ and nonzero $k_\perp$, reaching six at $\beta=2$. Equation~\eqref{eq:s_vanishing} states the leading contributions across regimes; its smaller displayed term need not be the complete next-order expansion on an individual path.

For a finite observation window and colored noise, let $\mathcal Q$ project off the independent constant and linear EP cluster columns and the allowed ordinary-response amplitude and frequency tangents, namely the waveform derivatives with respect to those fit parameters. With dimensionless center $s_0$,
\begin{equation}
    d_\EP^2=\frac{\epsilon^4}{4}\min_a
    \norm{\mathcal Q[e^{s_0x}x^2r(a)]}_N^2+o(\epsilon^4).
    \label{eq:s_weighted}
\end{equation}
For a real readout, take the real part inside the projection and resolve complex coefficients into real quadratures. The rank of the retained tangent space is assumed constant in the neighborhood considered. The common-center mismatch must therefore survive the physical readout and ordinary-response fitting to contribute observable information.

\subsection{Uniform limit near collinear excitation}
Decompose $k=\lambda c+\delta_\perp\norm{c}v$, with $c^\dagger v=0$, $\norm{v}=1$, and $\lambda=c^\dagger k/\norm{c}^2$. Equation~\eqref{eq:s_shared} becomes
\begin{equation}
    \begin{aligned}
        d_{\EP}^2&=\frac{\norm{c}^2\epsilon^4}{32}
        \min_a\left[|P(a)|^2+4\delta_\perp^2|a|^2\right]+O(\epsilon^5),\\
        P(a)&=a^2-2\lambda a+\zeta.
    \end{aligned}
    \label{eq:s_functional}
\end{equation}
For fixed $\Delta=\lambda^2-\zeta\ne0$, the two zero-mismatch centers at $\delta_\perp=0$ are $a_\pm=\lambda\pm\sqrt\Delta$. These stationary centers continue away from collinearity because the Hessian with respect to the real and imaginary parts of $a$ is positive definite at each simple root. Writing $a=a_0+u\delta_\perp^2+O(\delta_\perp^4)$, minimization gives
\begin{equation}
    \begin{aligned}
        u&=-\frac{a_0}{|\Delta|},\\
        \min_a\left[|P(a)|^2+4\delta_\perp^2|a|^2\right]
        &=4\mu\delta_\perp^2-\frac{4\mu}{|\Delta|}\delta_\perp^4
        +O(\delta_\perp^6),
    \end{aligned}
    \label{eq:s_root_continuation}
\end{equation}
where $\mu=\min(|a_+|^2,|a_-|^2)$. The denominator follows from $|P'(a_0)|^2=4|\Delta|$.

The collinear quadratic mismatch vanishes at $a_0$. The cubic mismatch there is
\begin{equation}
    \frac{\zeta\lambda-3a_0^2\lambda+2a_0^3}{6}
    =\frac{a_0\Delta}{3}.
    \label{eq:s_cubic_identity}
\end{equation}
The next center adjustment removes its component along $cg_2$, where $g_n=x^ne^{-x}$. The leading residual contributions are therefore
\begin{equation}
    -a_0\epsilon^2\delta_\perp\norm{c}vQ_{01}g_2,
    \qquad \frac{a_0\Delta}{3}\epsilon^3cQ_{012}g_3.
    \label{eq:s_two_residuals}
\end{equation}
Here $Q_{01}$ and $Q_{012}$ remove the listed lower moments. These contributions are orthogonal in record space. Using their moment norms in Eq.~\eqref{eq:s_polynomials} yields
\begin{equation}
    \begin{aligned}
        d_{\EP}^2&=\norm{c}^2\mu\left(
        \frac{\epsilon^4\delta_\perp^2}{8}
        +\frac{\epsilon^6|\Delta|^2}{32}\right)\\
        &\quad+o\left[\norm{c}^2(\epsilon^4\delta_\perp^2+\epsilon^6)\right].
    \end{aligned}
    \label{eq:s_uniform_vector}
\end{equation}
The coefficients in Eq.~(7) of the Letter are therefore $A=\norm{c}^2\mu/8$ and $B=\norm{c}^2\mu|\Delta|^2/32$. This expansion is uniform in the relative rate at which $\epsilon$ and $\delta_\perp$ approach zero, for fixed $c\ne0$ and $\Delta$ bounded away from zero. The remainder is controlled by $O(\epsilon^4\delta_\perp^4+\epsilon^5\delta_\perp^2+\epsilon^7)$. The distance is even in $\delta_\perp$ because its sign can be changed by a unitary transformation of the orthogonal record direction. Corrections to the matching center start at $\delta_\perp^2$, and the moment integrals converge uniformly in a stable neighborhood of the limiting center.

The minimization over all stable centers selects these local branches. At $\epsilon=0$, the nonzero single exponential is represented exactly only at $q=-1$. Closed sets of centers separated from this value have positive distance; centers approaching the stability boundary or infinity have vanishing or nonunit projection onto that exponential. An $o(\epsilon^4)$ candidate then requires $q+1=O(\epsilon)$ and $P(a)\to0$. Both branches in Eq.~\eqref{eq:s_two_residuals} have their quadratic and cubic coefficients multiplied by $|a_0|^2$, including the case in which the two magnitudes agree.

The observable crossover and required SNR are
\begin{equation}
    \begin{aligned}
        \Xi&=\frac{2\delta_\perp}{\epsilon|\Delta|},&
        \frac{32d_{\EP}^2}{\norm{c}^2\mu\epsilon^6|\Delta|^2}&\longrightarrow1+\Xi^2,\\
        \rho_{\mathrm{req}}&\sim
        \frac{4\sqrt{\Lambda_*}}{\sqrt\mu\,\epsilon^3|\Delta|\sqrt{1+\Xi^2}},&
        E_h&\longrightarrow\norm{c}^2/2.
    \end{aligned}
    \label{eq:s_snr_uniform}
\end{equation}
The last expression fixes the distance noncentrality $\rho^2d_{\EP}^2/E_h=\Lambda_*$. The scaling collapse in the Letter includes frequency splitting ($\zeta=-1$), damping splitting ($\zeta=1$), and a complex splitting direction. Finite-separation distances are calculated with Eq.~\eqref{eq:s_cauchy}.

\subsection{Excitation paths and pole cancellation}
For $j=\epsilon^\beta k$, the generic distance powers are collected in Table~\ref{tab:s_paths}. When $0<\beta<1$, the optimal displacement is $q+1=\epsilon^{2-\beta}\alpha$. Its leading coefficient is obtained by projecting $\zeta c$ off $k$:
\begin{equation}
    \alpha=\frac{\zeta k^\dagger c}{2\norm{k}^2},\qquad
    d_{\EP}^2\sim\frac{\epsilon^4}{32}\norm{c_{\perp k}}^2.
\end{equation}
For one record this term vanishes, leaving the single-response power $4+2\beta$. For $\beta>1$, replace $\lambda,\delta_\perp$ in Eq.~\eqref{eq:s_uniform_vector} by $\epsilon^{\beta-1}\lambda_0,\epsilon^{\beta-1}\delta_0$; the two orders are $2\beta+2$ and six. Finite $j$ uses the regular tangent information of the following sections.

\par\medskip\noindent
\begin{minipage}{\columnwidth}
    \captionof{table}{Leading powers of $d_{\EP}^2$ for generic nonvanishing $c,k$. The last column assumes noncollinear excitations.}
    \label{tab:s_paths}
    \begin{ruledtabular}
    \begin{tabular}{ccc}
        Path & Single record & Shared center\\
        $\beta=0$ & $\epsilon^4$ & $\epsilon^4$\\
        $0<\beta<1$ & $\epsilon^{4+2\beta}$ & $\epsilon^4$\\
        $\beta=1$ & $\epsilon^6$ & $\epsilon^4$\\
        $1<\beta<2$ & $\epsilon^6$ & $\epsilon^{2\beta+2}$\\
        $\beta\ge2$ & $\epsilon^6$ & $\epsilon^6$
    \end{tabular}
    \end{ruledtabular}
\end{minipage}
\par\medskip
Along a smooth physical parameter $p$ with $\chi\propto p$, $J_0\ne0$ gives $\beta=0$, whereas $J_0=0,J_1\ne0$ gives $\beta=2$. A source controlled as $\sqrt{|p|}$ realizes $\beta=1$. These choices concern how a fixed spectral neighborhood is excited.

\begin{figure*}[t]
    \centering
    \subfloat[]{\includegraphics[width=.47\textwidth]{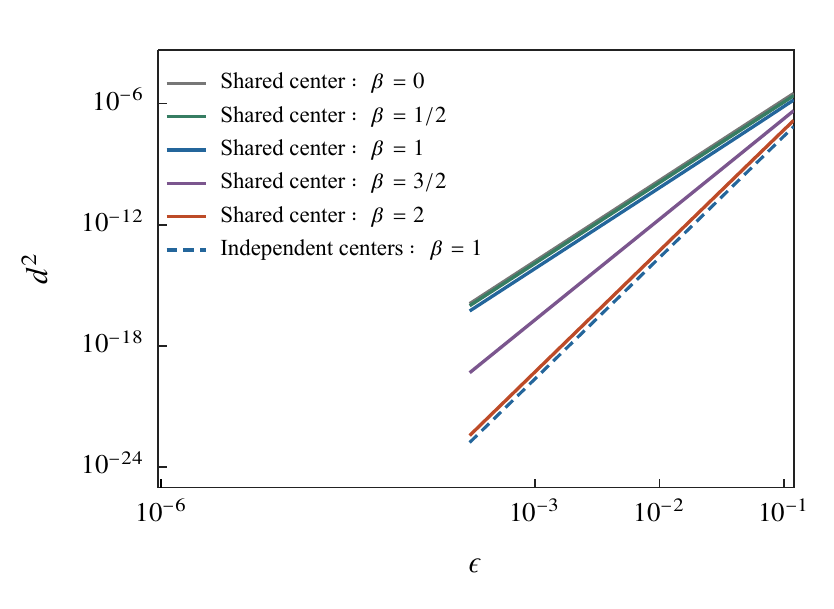}}
    \hfill
    \subfloat[]{\includegraphics[width=.47\textwidth]{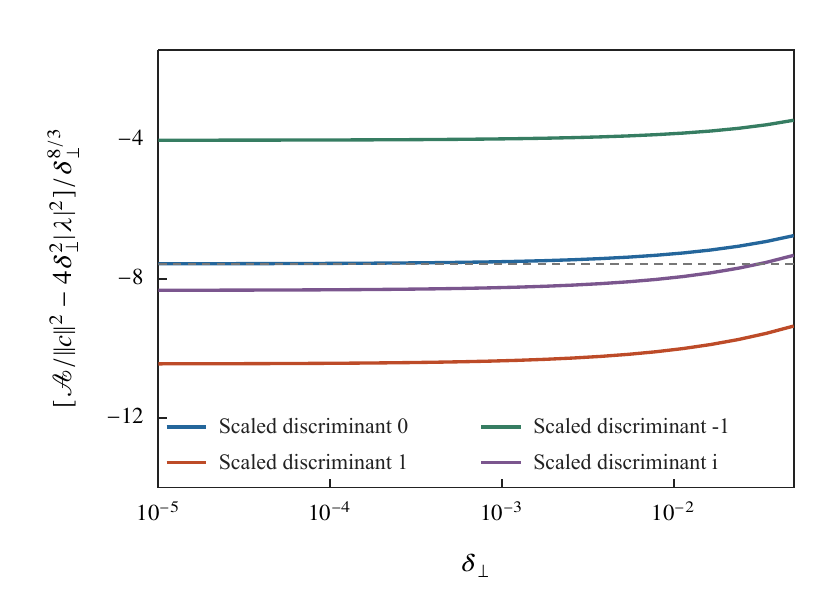}}
    \caption{(a) Exact shared-center distances for $c=(1,1)/\sqrt2$, $k=(0.9,0.1)/\sqrt2$, $\zeta=-1$. For $\beta=1$, the dashed curve uses separately fitted EP centers for the same two records. (b) The first boundary correction to the fourth-order coefficient with $\Delta=\delta_\perp^{4/3}\mathfrak D$, $\lambda=\sqrt{1+\Delta}$ and $\zeta=1$. The dashed line at $\mathfrak D=0$ is $-6\,2^{1/3}$.}
    \label{fig:s_paths_boundary}
\end{figure*}

At $\Delta=0$ and $\delta_\perp=0$, one residue vanishes and the response is a single exponential. The matching roots then coincide. With $a=\lambda(1-t)$ and $|\lambda|=1$, the quantity minimized in Eq.~\eqref{eq:s_functional} is $t^4+4\delta_\perp^2(1-t)^2$. Its minimum satisfies
\begin{equation}
    \begin{aligned}
        t^3&=2\delta_\perp^2(1-t),\qquad
        t\sim2^{1/3}\delta_\perp^{2/3},\\
        \min_a\left[|P(a)|^2+4\delta_\perp^2|a|^2\right]
        &=4\delta_\perp^2-6\,2^{1/3}\delta_\perp^{8/3}\\
        &\quad+4\,2^{2/3}\delta_\perp^{10/3}+O(\delta_\perp^4).
    \end{aligned}
    \label{eq:s_cancel}
\end{equation}
For $\Delta=\delta_\perp^{4/3}\mathfrak D$ and $a-\lambda=\delta_\perp^{2/3}u$, the same quantity becomes
\begin{equation}
    \begin{aligned}
        4\delta_\perp^2|\lambda|^2
        &+\delta_\perp^{8/3}\left[|u^2-\mathfrak D|^2
        +8\Ree(\bar\lambda u)\right]\\
        &+4\delta_\perp^{10/3}|u|^2.
    \end{aligned}
    \label{eq:s_cancel_layer}
\end{equation}
This gives the boundary layer of the fourth-order distance coefficient when the two center branches merge. Figure~\ref{fig:s_paths_boundary} shows excitation paths and the coefficient near this boundary.

\section{Radiative channels and independent center information}
For two polarization readouts with antenna matrix $F$, the excitation determinant transforms as
\begin{equation}
    \det[c_{\mathrm{det}},j_{\mathrm{det}}]
    =\det(F)(c_+j_\times-c_\times j_+).
    \label{eq:s_antenna_det}
\end{equation}
The radiation must supply distinct relative excitations, and the readout must retain them. This condition concerns the physical source as well as the number of detectors.

\subsection{One Jordan block and degenerate radiative channels}
In dimensionless time, write a single block in the regular form
\begin{equation}
    D=\begin{pmatrix}0&g\\ \ch/g&0\end{pmatrix},\qquad
    c=Rb,\quad j=RDb,\quad b=(b_1,b_2)^T,
\end{equation}
where $g$ has a finite nonzero limit. For a fixed invertible readout $R$,
\begin{equation}
    \det[c,j]=\det(R)\left(\frac{\ch b_1^2}{g}-gb_2^2\right).
    \label{eq:s_single_block}
\end{equation}
If $R$ and its inverse remain bounded, $c$ has a nonzero limit and $j=O(\epsilon)$ imply $b_2=O(\epsilon)$. Thus $\det[c,j]=O(\epsilon^2)$ and $\delta_\perp=O(\epsilon)$. The same conclusion holds for additional fixed full-rank readouts with a bounded left inverse. A weakly excited single block generically retains the sixth-order distance even when both state directions are measured.

For $g=1$, $\zeta=-1$, $R=I$, and real $\lambda_0$, the explicit path $b=(1,\epsilon\lambda_0)^T$ gives
\begin{equation}
    \begin{aligned}
        c&=(1,\epsilon\lambda_0)^T,&
        j&=(\epsilon\lambda_0,-\epsilon^2)^T,\\
        \delta_\perp&=\frac{\epsilon(1+\lambda_0^2)}{1+\epsilon^2\lambda_0^2},&
        \Xi&\longrightarrow2.
    \end{aligned}
    \label{eq:s_single_path}
\end{equation}
Two independently excited degenerate radiative channels instead have $D_{\mathrm{tot}}=I_2\otimes D$. Their minimal polynomial remains quadratic, and the EP contains two Jordan blocks of order two. The initial state of channel $m$ can be $(c_m,\epsilon k_m/g)^T$; reading its first component yields $c_m,j_m=\epsilon k_m$. Independent excitations of these radiative channels can therefore keep $\delta_\perp$ finite while preserving a common center and splitting.

Axial gravitational perturbations of a spherical background provide such angular channels. Different $m$ at fixed $\ell$ obey the same radial equation. In the Cunningham--Price--Moncrief (CPM) normalization, the asymptotic odd-parity strain is proportional to \cite{MartelPoisson2005}
\begin{equation}
    h_+-\ii h_\times\ \propto\ \frac{\ii}{r}
    \sum_m\Psi^{\mathrm{odd}}_{\ell m}\,{}_{-2}Y_{\ell m}.
    \label{eq:s_cpm_strain}
\end{equation}
For $\ell=2$, $m=\pm2$, at azimuth zero, write $\Psi_{22}=X+\ii Y$ and use the complex-conjugation relation required by a real metric. With
\begin{equation}
    \mathcal Y_\pm(\iota)=\sqrt{\frac5{64\pi}}(1\pm\cos\iota)^2,
\end{equation}
the strain is proportional to $\ii[(\mathcal Y_++\mathcal Y_-)X+\ii(\mathcal Y_+-\mathcal Y_-)Y]$. Two independent real radial excitations $X,Y$ map to distinct polarizations when $\cos\iota\ne0$. A specified merger model can correlate these initial data; those correlations enter through its allowed excitation family. Figure~\ref{fig:s_radiation}(a) compares a proportional response, Eq.~\eqref{eq:s_single_path}, and independently excited degenerate radiative channels in the same strain likelihood. The corresponding small-splitting squared-distance slopes are $6.0004$, $6.0006$, and $4.0012$.

\subsection{A finite-precision center measurement}
\begin{figure*}[t]
    \centering
    \subfloat[]{\includegraphics[width=.47\textwidth]{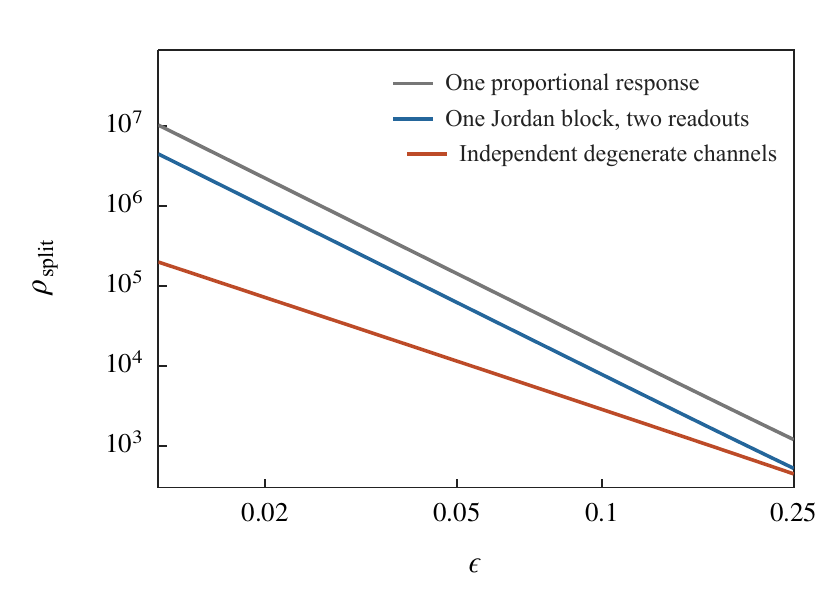}}
    \hfill
    \subfloat[]{\includegraphics[width=.47\textwidth]{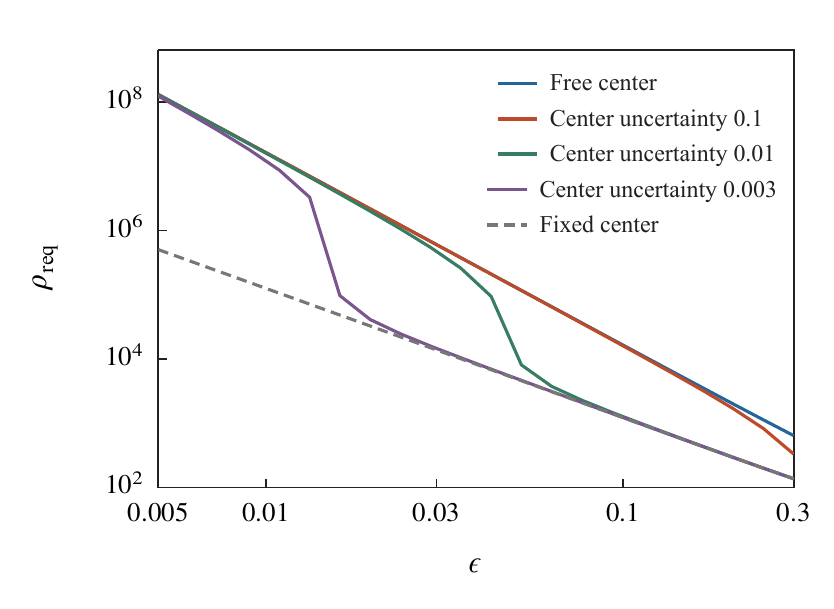}}
    \caption{(a) Resolving the same spectral separation through different radiative structures. One weakly excited Jordan block with two fixed readouts retains the sixth-order distance; independently excited degenerate radiative channels retain fourth order. The strain calculation uses $f_c=150$ Hz, $Q=3$, and aLIGO design noise. (b) Collinear white half-line response with $c=1$, $\lambda=1/2$, $\zeta=-1$: free, finite-precision, and fixed centers at joint distance noncentrality 10. A fixed finite $\sigma_q$ gives a finite-separation improvement.}
    \label{fig:s_radiation}
\end{figure*}

Let an independent measurement give the dimensionless center $q_0$, with variance $\sigma_q^2$ in each real component. The target center is $q_0$ and the candidate EP has $q=q_0+\epsilon a$. With waveform noise scale $\sigma_w$, its noise-free joint likelihood penalty is
\begin{equation}
    \frac{d^2(q)}{\sigma_w^2}+\frac{\epsilon^2|a|^2}{\sigma_q^2}.
\end{equation}
Factoring the leading coefficient in Eq.~\eqref{eq:s_functional} gives
\begin{equation}
    \begin{aligned}
        \frac{\norm{c}^2\epsilon^4}{32\sigma_w^2}
        \min_a\left\{|P(a)|^2+(4\delta_\perp^2+\kappa)|a|^2\right\},\\
        \kappa=\frac{32\sigma_w^2}{\norm{c}^2\epsilon^2\sigma_q^2}
        =\frac{32E_h}{\norm{c}^2\rho^2\epsilon^2\sigma_q^2}.
    \end{aligned}
    \label{eq:s_center_functional}
\end{equation}
Independent excitation and center information thus contribute through $\delta_{\mathrm{eff}}^2=\delta_\perp^2+\kappa/4$ at this order. For collinear excitation, $\Phi(\kappa)=\min_a[|P(a)|^2+\kappa|a|^2]$ has limits
\begin{equation}
    \Phi(\kappa)=\mu\kappa-\frac{\mu\kappa^2}{4|\Delta|}+O(\kappa^3),
    \qquad \Phi(\infty)=1.
\end{equation}
The second limit fixes the center and restores the quadratic residual. For $c=1$, real $\lambda$, and $\zeta=-1$, the fixed-center expansion is
\begin{equation}
    d_{q=-1}^2=\frac{\epsilon^4}{32}+\frac{3\lambda\epsilon^5}{32}
    +\frac{(5\lambda^2-6)\epsilon^6}{64}+O(\epsilon^7).
\end{equation}
At a fixed finite $\sigma_q$, resolution improves over a finite separation interval. Along the freely centered resolution threshold $\rho\sim\epsilon^{-3}$, $\kappa\sim\epsilon^4$, and the asymptotic exponent returns to the free-center value. Figure~\ref{fig:s_radiation}(b) uses the full joint likelihood penalty at finite separation, including this crossover. Sustained quartic resolution requires the center uncertainty to decrease sufficiently to constrain the competing $O(\epsilon)$ center displacement.

An event can supply center information through other resonances or through remnant parameters inferred from earlier radiation. This inference uses a theory-dependent relation between remnant parameters and the EP cluster center. Correlated parts of the same data enter one joint likelihood. The noise trials below use a statistically independent center measurement, with both real components randomized.

\subsection{Center information with additional ordinary responses}
The controls supporting Sec.~VI of the Letter use the tensor signal and finite-window noise model of Sec.~XI, at $f_c=150$ Hz, $Q=3$, $\epsilon=0.16$, and $\rho=10^4$. There are two true ordinary components, with dimensionless frequencies $-2.1+7.4\ii$ and $-3+8.2\ii$, amplitudes $A_o$ from Eq.~\eqref{eq:s_strain_ordinary} and $0.015e^{0.4\ii}$, and polarization direction $c=(1,1)/\sqrt2$. The additional candidate frequencies are $-4+9\ii$, $-5+9.8\ii$, and $-6+10.6\ii$, with zero injected amplitudes. Their fitted damping ranges are 0.7--1.4 times these reference values, and their oscillation frequencies vary by $\pm0.35$. All retained frequencies and amplitudes are profiled.

Let $u=(\Ree q,\Imm q)^T$ denote the center in units of the true decay rate. Independent Gaussian information adds $(u-u_{\rm obs})^T\Sigma_u^{-1}(u-u_{\rm obs})$ to the whitened residual square. Table~\ref{tab:s_multimode_center} sets $u_{\rm obs}=(-1,6)^T$ and $\Sigma_u=\sigma_q^2I$. These are noise-free likelihood differences; noisy external measurements are treated separately in Sec.~XI. The two orthogonal antenna rows and proportional excitation are held fixed. Finite center precision preserves a substantial gain for each of the tested numbers of ordinary response terms.

\begin{table}[htbp]
    \caption{Noise-free $T_\chi$ with two true ordinary responses and $N_{\rm ord}$ fitted ordinary response terms. The center errors are dimensionless.}
    \label{tab:s_multimode_center}
    \begin{ruledtabular}
    \begin{tabular}{crrr}
    $N_{\rm ord}$ & Free center & $\sigma_q=0.01$ & Fixed center\\
    2 & 3.104 & 172.05 & 258.97\\
    3 & 1.523 & 108.14 & 110.11\\
    5 & 0.709 & 32.80 & 33.25
    \end{tabular}
    \end{ruledtabular}
\end{table}

The relevant measurement direction follows from the compensating center shift. For $\lambda=1/2$ and $\zeta=-1$, the matching roots are $a_\pm=(1\pm\sqrt5)/2$, both real. The leading compensation therefore changes the decay rate, so measuring that direction is especially useful. A complex splitting direction generally requires the corresponding combination of frequency and decay-rate information.

The same two-target comparison with distinct excitation, $\delta_\perp=0.4$, also isolates the shared-center constraint while keeping the ordinary frequencies shared. With two fitted ordinary response terms, separate and shared EP centers give $T_\chi=4.297$ and $62.65$; with five they give $0.491$ and $10.64$. This extends the center-sharing mechanism to a finite window with ordinary responses and detector noise weighting.

\section{Finite Jordan information and physical resources}
For fixed $j\ne0$, the derivative with respect to $\ch$ at the EP is $e^{-x}(cx^2/2+jx^3/6)$. After independent amplitudes and the common center are fitted, its information for white noise at positive times is
\begin{equation}
    \Gamma=\frac1{\sigma_x^2}\left(\frac{\norm{c_\perp}^2}{32}
    +\frac{\norm{j}^2}{128}\right),\qquad
    c_\perp=c-j\frac{j^\dagger c}{\norm{j}^2},
    \label{eq:s_finite_info}
\end{equation}
where a complex response is resolved into its real and imaginary quadratures, each with variance density $\sigma_x^2$. The center tangent removes the quadratic component parallel to $j$, leaving the transverse excitation term and the cubic term. Related information matrices for multiple poles in colored noise are discussed in Ref.~\cite{Badeau2008}.

The EP signal energy obeys
\begin{equation}
    \begin{aligned}
        E_h&=\tfrac12\norm{c+j/2}^2+\tfrac18\norm{j}^2,\\
        E_h-16\sigma_x^2\Gamma
        &=\frac{\norm{j}^2}{2}
        \left|\frac{j^\dagger c}{\norm{j}^2}+\frac12\right|^2\ge0.
    \end{aligned}
    \label{eq:s_resource}
\end{equation}
An unrestricted single response with $c=-j/2$ saturates the bound. The advantage of multiple physical excitations consequently depends on the accessible source space, its propagation, and the fitted ordinary responses.

For real initial-data vectors $w_j$, introduce $Q=\sum_jw_jw_j^T\succeq0$. Initial energy and whitened output power are $\operatorname{Tr}(EQ)$ and $\operatorname{Tr}(SQ)$, respectively. After record-specific amplitudes are removed, the shared-parameter Gram matrix $G(Q)$ is linear in $Q$. Profiling over center and ordinary-frequency parameters $\nu$ gives
\begin{equation}
    F(Q)=G_{\chi\chi}-G_{\chi\nu}G_{\nu\nu}^{+}G_{\nu\chi}.
\end{equation}
We maximize $\lambda_{\min}F(Q)$ subject to $\operatorname{Tr}(EQ)=1$. A conditional-gradient iteration uses a supporting upper bound for this concave objective. The achieved value is $8.1378\times10^{-9}$ and the bound is $8.1678\times10^{-9}$. The two principal covariance directions, renormalized to unit total initial energy, define the two-source experiment in the Letter.

For comparison at finite splitting, we evaluate 110 unit-energy single-source candidates and select the one with the largest profiled distance to $|\ch|\ge0.0045$. This is a finite-search comparison within the specified source basis. Source selection uses the $\Delta x=0.05,0.025$ extrapolation; reported distances use $0.025,0.0125$. The squared distances and output powers are
\begin{equation}
    \begin{array}{c|cc}
        &d_{\rm ext}^2&\sum_j\norm{W\psi_j}^2\\\hline
        \text{two sources}&2.88679\times10^{-14}&1.93731\times10^{-4}\\
        \text{selected single}&1.01118\times10^{-15}&3.38551\times10^{-4}.
    \end{array}
    \label{eq:s_distances}
\end{equation}
The distance ratio is $28.55$, and its square root is $5.34$. Two proportional copies divided by $\sqrt2$ have the same total energy and the same complete profile distance as their original single response. Changes in overall amplitude therefore do not reproduce excitation diversity.

\section{Einstein--Maxwell--scalar evolution}
The background action and metric are
\begin{equation}
    \begin{aligned}
        S&=\int\dd^4x\sqrt{-g}\,[R-2(\partial\phi)^2-e^{\alpha\phi^2}F^2],\\
        \dd s^2&=-Ne^{-2d}\dd t^2+N^{-1}\dd r^2+r^2\dd\Omega^2.
    \end{aligned}
\end{equation}
Eliminating the electric field gives the background equations \cite{Cheng2026}
\begin{equation}
    \begin{aligned}
        N'&=\frac{1-N}{r}-\frac{Q^2}{r^3e^{\alpha\phi^2}}-rN(\phi')^2,\\
        (r^2N\phi')'&=-\frac{\alpha Q^2\phi}{r^2e^{\alpha\phi^2}}-r^3N(\phi')^3,\\
        d'&=-r(\phi')^2.
    \end{aligned}
\end{equation}
Set $r_H=1$, use the regular horizon expansion, shoot in $(Q,\phi_H)$ for $\phi_\infty=0$ and the desired $q_{\mathrm{BH}}=Q/M$, and normalize $d(\infty)=0$. The horizon coefficients are $N'_H=1-Q^2e^{-\alpha\phi_H^2}$ and $\phi'_H=-\alpha Q^2\phi_He^{-\alpha\phi_H^2}/N'_H$. At the nominal EP, $Q=3.796409062492$, $\phi_H=2.203020740145$, and $M=3.642718583726$, giving $q_{\mathrm{BH}}=1.0421911479663$. Seven backgrounds have $q_{\mathrm{BH}}-q_\EP=0,\pm10^{-4},\pm2\times10^{-4},\pm4\times10^{-4}$.

For the neutral test scalar $\Phi=r^{-1}\psi(t,r)Y_{\ell m}$, define $x=r_*/M$, $\bar t=t/M$, and $\dd r_*/\dd r=e^d/N$. Then
\begin{equation}
    \begin{aligned}
        \partial_{\bar t}^2\psi&=\partial_x^2\psi-\mathcal V\psi,\\
        \mathcal V&=M^2\frac{Ne^{-2d}}{r^2}
        \left[\ell(\ell+1)+1-N-\frac{Q^2}{r^2e^{\alpha\phi^2}}\right].
    \end{aligned}
    \label{eq:s_pde}
\end{equation}
We use $\ell=2$ and place the potential maximum at $x=0$. Shape-preserving cubic interpolation gives the potential. At the near-horizon end of the background table, an exponential continuation uses the local logarithmic slope, approaching $MN'_He^{-d_H}$. The outer background covers the entire evolution domain.

Initial field and velocity are $f(x),g(x)$. For $G_{p,w}=\exp[-(x-p)^2/(2w^2)]$, choose $(p,w)=(-15,2),(-7.5,2),(0,2),(7.5,2),(15,2),(0,6)$, and place each profile in either $f$ or $g$. The basis order alternates field and velocity at each listed position. Each initial profile is normalized using
\begin{equation}
    E[f,g]=\frac12\int\big[g^2+(\partial_xf)^2+\mathcal V f^2\big]\dd x.
    \label{eq:s_energy}
\end{equation}
The full off-diagonal Gram matrix is retained for source superpositions. Unit energy is a reference normalization for these linear solutions.

On $-320\le x\le320$, we use centered second-order space and time differences with $\Delta\bar t=\Delta x/2$. Writing $L_h f=D_{xx}f-\mathcal V f$, the first step is
\begin{equation}
    \psi^1=f+\Delta\bar t\,g
    +\tfrac12\Delta\bar t^2L_h f+\tfrac16\Delta\bar t^3L_h g,
\end{equation}
followed by $\psi^{n+1}=2\psi^n-\psi^{n-1}+\Delta\bar t^2L_h\psi^n$. Outputs at $x_o=2,30,60$ are sampled every $0.2M$ up to $\bar t=230$. Their domains of dependence remain inside the computational interval. Grids $\Delta x=0.05,0.025,0.0125$ are available for all seven backgrounds, with $0.1$ also used at $0,\pm2\times10^{-4}$. The adopted waveform is $(4\psi_{0.0125}-\psi_{0.025})/3$. This removes the leading second-order propagation error. The retarded coordinate is $u/M=\bar t-x_o$.

For source control, let $K$ map the twelve real coefficients to $(\Ree J,\Imm J)$ at the EP. The energy-orthogonal projector is
\begin{equation}
    P_J=E^{-1}K^T(KE^{-1}K^T)^{-1}K.
\end{equation}
Normalize $(I-P_J)w$ and $P_Jw$ to obtain $w_0,w_1$. The family $(w_0+\eta w_1)/\sqrt{1+\eta^2}$ has unit initial energy. At $x_o=60$ and $u_0/M=70$, $\eta=0.001,0.01,0.1,1$ yields $|J/(\gamma C)|=0.00171,0.01835,0.2983,0.1960$. The nonmonotonic ratio reflects the accompanying change in $C$. The suppression floor varies to about $1.1\times10^{-4}$ under the readout and window changes used here.

Two independently constructed weak-source families, with $\eta=\sqrt{|\ch|}$, give excitation misalignment $1-|c^\dagger k|^2/(\norm{c}^2\norm{k}^2)=0.132$--$0.612$ across the six neighboring backgrounds. Their fourth-order coefficient in Eq.~\eqref{eq:s_shared}, normalized by $E_0=\norm{c}^2/2$, is $0.0108$--$0.0392$. These finite-separation measurements characterize the physical sources relative to the common-center condition; the asymptotic distance curves are calculated from Eq.~\eqref{eq:s_cauchy}.

\par\medskip\noindent
\begin{minipage}{\columnwidth}
 \centering
 \includegraphics[width=\columnwidth]{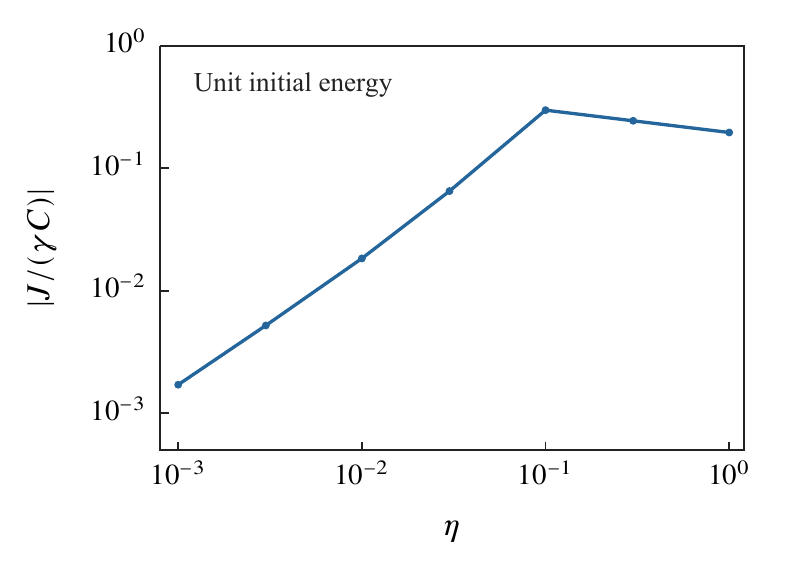}
 \captionof{figure}{Jordan excitation controlled by real initial data at fixed BH background and unit initial energy. The reference is $u_0/M=70$ at $x_o=60$.}
\end{minipage}
\par\medskip

\subsection{Parameter precision and the observable EP neighborhood}
This subsection supplies the parameter and inference controls for Sec.~IV of the Letter. The spectral reference is $p_*=(\alpha_*,q_*)=(0.79040752752,1.04219114847)$. Independent spectral calculations near that point give the local map
\begin{equation}
    \begin{pmatrix}\Ree\ch\\\Imm\ch\end{pmatrix}
        \simeq D\,\Delta p,\qquad
        D=\begin{pmatrix}5.13482&-9.91079\\5.95883&-62.80765\end{pmatrix}.
        \label{eq:s_parameter_map}
    \end{equation}
Thus $|\ch|<r$ gives the local ellipse $\Delta p^TD^TD\Delta p<r^2$. The singular values of $D$ are 63.9368 and 4.12046; simultaneous errors in the two parameters have different tolerances along different directions. Increasing spectral order from 100 to 120 changes $D$ by $2.4\times10^{-5}$ in relative matrix norm. At $r=0.005$, the directly calculated nonlinear boundary differs radially from the ellipse by at most 1.49\%. At fixed $\alpha$, this boundary is $-7.9208\times10^{-5}<q-q_*<7.8067\times10^{-5}$. At fixed $q$, it is $-6.2699\times10^{-4}<\alpha-\alpha_*<6.4451\times10^{-4}$.

Table~\ref{tab:s_rounding} evaluates the spectrum after rounding both reference parameters. Five decimal places suffice for displaying this example within the allowance $b_\chi=5\times10^{-4}$. The calculations retain full input precision. For a measurement, the reported parameter uncertainty determines meaningful digits; the spectral ellipse maps that uncertainty into a splitting scale.
\begin{table}[htbp]
    \caption{Spectral splitting after rounding both $\alpha_*$ and $q_*$.}
    \label{tab:s_rounding}
    \begin{ruledtabular}
    \begin{tabular}{cr}
    Decimal places & $|\ch|$\\
    3 & $9.5010\times10^{-3}$\\
    4 & $6.1454\times10^{-4}$\\
    5 & $9.0125\times10^{-5}$\\
    6 & $1.2750\times10^{-5}$
    \end{tabular}
    \end{ruledtabular}
\end{table}

The near-EP waveform experiments use the nominal evolution inputs $\alpha=0.79040751953125$ and $q_0=1.042191148$, with the four offsets in Table~\ref{tab:s_near_ep}. The small difference between this nominal point and the spectral reference is retained in computing the true splitting. A fixed pair of source configurations, normalized to unit total energy on each background, is evolved and fitted on $70\le u/M\le150$ at $x_o=60$, using two ordinary components, four tail terms, and the correlated Gaussian noise of Sec.~VII. Noise-free waveform recovery errors are $1.14$--$1.25\times10^{-4}$ in $\ch$ at these four backgrounds, below the adopted allowance. Each SNR and background has 200 trials; comparisons use paired noise. The thresholds are $t_J=15$, $t_\chi=10$, and $T_{\rm res}(0.0045)>10$.

\begin{table}[htbp]
    \caption{Near-EP classification, as counts (E,R) out of 200. E denotes the full finite-resolution EP criterion; R denotes resolved nonzero splitting with the same upper bound. Remaining outcomes satisfy neither class.}
    \label{tab:s_near_ep}
    \setlength{\tabcolsep}{3pt}
    \begin{ruledtabular}
    \begin{tabular}{crcc}
    $10^6(q-q_0)$ & True $|\ch|$ & $\rho=4\times10^5$ & $\rho=10^6$\\
    8  & 0.000509 & (190,4) & (121,79)\\
    16 & 0.001019 & (147,46) & (3,197)\\
    32 & 0.002041 & (51,137) & (0,200)\\
    64 & 0.004094 & (0,3) & (0,116)
    \end{tabular}
    \end{ruledtabular}
\end{table}
The true spectrum lies within $|\ch|<0.005$ at all four points. The observed upper bound is harder to establish near its edge, where statistical uncertainty and the allowance consume the remaining margin. For fixed nonzero splitting, increasing SNR can instead exclude zero while retaining the upper bound. The numerical allowance is supported by the specified waveform controls and these tested backgrounds.

\begin{figure}[t]
    \centering
    \includegraphics[width=\columnwidth]{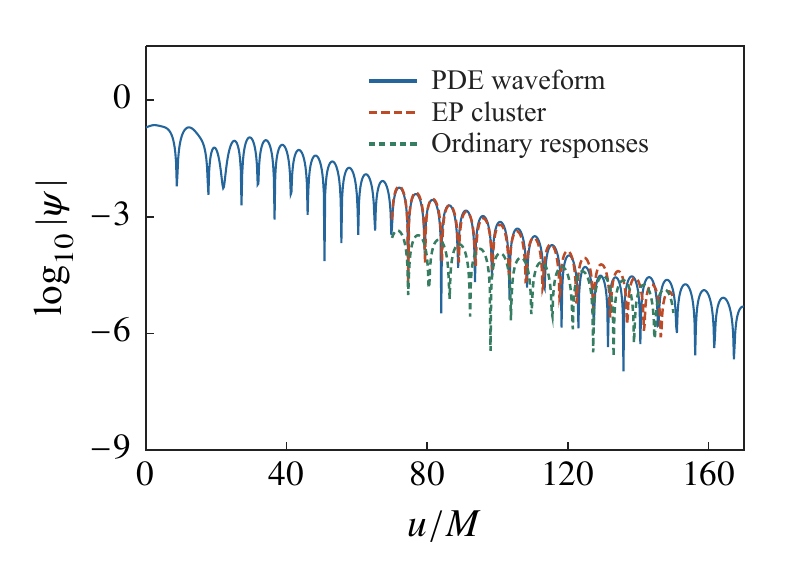}
    \caption{The second member of the source pair with unit total energy at the nominal EMS EP, with the readout and time window of Fig.~1(a) of the Letter. The common spectrum permits different EP cluster and ordinary amplitudes.}
    \label{fig:s_second_waveform}
\end{figure}

\section{Finite-window response and likelihood}
All fits use the signed real field. On a window beginning at $u_0$, the model is
\begin{equation}
    \begin{aligned}
        \psi_j(\tau)&=\Ree\left[h_j(\tau)+\sum_{b=1}^{2}A_{jb}e^{s_b\tau}\right]\\
        &\quad+\sum_{p=7}^{10}b_{jp}
        \left(1+\frac{\tau}{u_0+r_{*o}}\right)^{-p},\qquad \tau=u-u_0.
    \end{aligned}
    \label{eq:s_model}
\end{equation}
Times and frequencies in the implementation are in units of $M$ and $M^{-1}$. The tail basis describes the late-time response on the fitted window; the leading power seven is appropriate to the $\ell=2$ scalar at fixed radius \cite{GundlachPricePullin1994}. A fifth term with $p=11$ gives the higher-order model. The two ordinary complex frequencies describe the slow oscillatory component and a faster finite-window component. The latter is used as an empirical damped response.

The shared EP cluster center has $\gamma M\in[0.035,0.18]$ and $\Omega M\in[0.58,0.73]$. Ordinary damping and frequency bands are $[0.025,0.075]\times[0.50,0.58]$ and $[0.12,0.32]\times[0.63,0.90]$. Representative EP-window centers are $Ms_1\simeq-0.04582+0.53836\ii$ and $Ms_2\simeq-0.183+0.724\ii$. The real and imaginary parts of $\ch$ range over $[-0.7,0.7]$ in the unconstrained split fit. The exterior profile searches splitting radii from the specified target to $0.7\sqrt{2}$, covering the entire square parameter domain with its circumscribed disk. Minimization over this disk gives a conservative exterior statistic. All spectral parameters are estimated from the data with independent amplitudes for each record.

For adjacent-sample correlation $r_n=0.6$, write $\operatorname{Cov}(n_i,n_k)=\sigma^2r_n^{|i-k|}$. The stationary whitening operation, which converts the correlated samples into samples with equal, uncorrelated noise, is
\begin{equation}
    (Wy)_0=y_0,\qquad
    (Wy)_i=\frac{y_i-r_n y_{i-1}}{\sqrt{1-r_n^2}},\quad i\ge1.
\end{equation}
Distinct records have independent noise. The main window contains 401 samples, with $u/M=70,70.2,\ldots,150$. Data and every design-matrix column undergo the same whitening. For a complex column $f$, the real columns are $(\Ree f,-\Imm f)$, so $\Ree(af)=a_R\Ree f-a_I\Imm f$.

At fixed nonlinear parameters $\vartheta$, set $A=WX/\sigma$, $d=Wy/\sigma$, and profile all linear coefficients using a column-scaled singular-value decomposition:
\begin{equation}
    \mathrm{RSS}(\vartheta)=\norm{(I-AA^+)d}^2,
    \qquad \ell_p=-\tfrac12\mathrm{RSS}+\mathrm{constant}.
\end{equation}
Variable projection eliminates the linear amplitudes at each nonlinear parameter point. The nonlinear fit uses the full derivative of this profiled residual, including its residual-dependent term. Near the EP the basis and derivatives use the globally convergent power series
\begin{equation}
    \CC_\chi=\sum_{n\ge0}\frac{\chi^n\tau^{2n}}{(2n)!},\qquad
    \SSS_\chi=\sum_{n\ge0}\frac{\chi^n\tau^{2n+1}}{(2n+1)!}.
\end{equation}
This keeps the same parameterization on both sides of zero splitting. Fits are initialized from multiple stable centers and ordinary frequencies.

Table~\ref{tab:s_scan} gives the seven-background sequence in the Letter. At the nominal EP the two-source $u_0/M=70$ fit gives $\ch=-8.23\times10^{-5}+9.66\times10^{-5}\ii$ and relative waveform residual $5.64\times10^{-7}$. At $u_0/M=80$, these become $-2.43\times10^{-5}-6.93\times10^{-6}\ii$ and $1.39\times10^{-7}$.
\par\medskip\noindent
\begin{minipage}{\columnwidth}
 \captionof{table}{Two-source waveform splitting at $x_o=60$, $80\le u/M\le160$.}
 \label{tab:s_scan}
 \begin{ruledtabular}\begin{tabular}{rrr}
 $10^4(q-q_\EP)$&$10^3\Ree\ch$&$10^3\Imm\ch$\\\hline
 $-4$&$3.7453$&$24.2236$\\
 $-2$&$1.9006$&$12.3204$\\
 $-1$&$0.9496$&$6.2120$\\
 $0$&$-0.0243$&$-0.0069$\\
 $1$&$-1.0241$&$-6.3414$\\
 $2$&$-2.0530$&$-12.7968$\\
 $4$&$-4.2115$&$-26.0934$\\
 \end{tabular}\end{ruledtabular}
\end{minipage}
\par\medskip

\section{Controlled weak excitation in complete waveforms}
We add four EMS backgrounds with $Q/M-q_{\EP}=8\times10^{-6},1.6\times10^{-5},3.2\times10^{-5},6.4\times10^{-5}$. Each background is obtained by shooting the field equations; interpolation supplies the shooting initial guess. The same twelve real initial-data profiles are evolved on $\Delta x=0.05,0.025,0.0125,0.00625$ with $\Delta\bar t=\Delta x/2$, on $[-280,280]$ up to $\bar t=212$. The observation windows lie within the domain unaffected by the boundaries.

Let $T$ map real source coefficients into $(\Ree C,\Imm C,\Ree j,\Imm j)$ at the selected spectral parameters, and let $E$ be the initial-energy matrix. For target coefficient columns $S$, the minimum-energy sources are
\begin{equation}
    W=E^{-1}T^T(TE^{-1}T^T)^{-1}S.
    \label{eq:s_source_design}
\end{equation}
The pair is then normalized to unit total energy. We use $c=(1,1)/\sqrt2$, $k=(0.9,0.1)/\sqrt2$, and prescribe $j=\epsilon k$ or $j=\epsilon^2k$. These initial data change the relative EP cluster excitation through the physical source map.

The final comparison uses $x_o=60$, $80\le u/M\le140$, two ordinary response terms with shared free frequencies, and six tail terms, $p=7,\ldots,12$. At each background the same two waveforms are fitted with EP models using a shared center or separately fitted centers. The ordinary frequencies remain shared in both cases. Define
\begin{equation}
    \mathcal D^2=\frac{\min\norm{W_N(y-m)}^2}{\norm{W_Ny}^2},
\end{equation}
where $W_N$ whitens the AR(1) covariance used above. This normalized squared fit residual compares a model with the complete waveform and includes error from the finite-window response model. The complete split fit supplies a residual baseline for interpreting the EP mismatch.

\begin{figure*}[t]
    \centering
    \subfloat[]{\includegraphics[width=.47\textwidth]{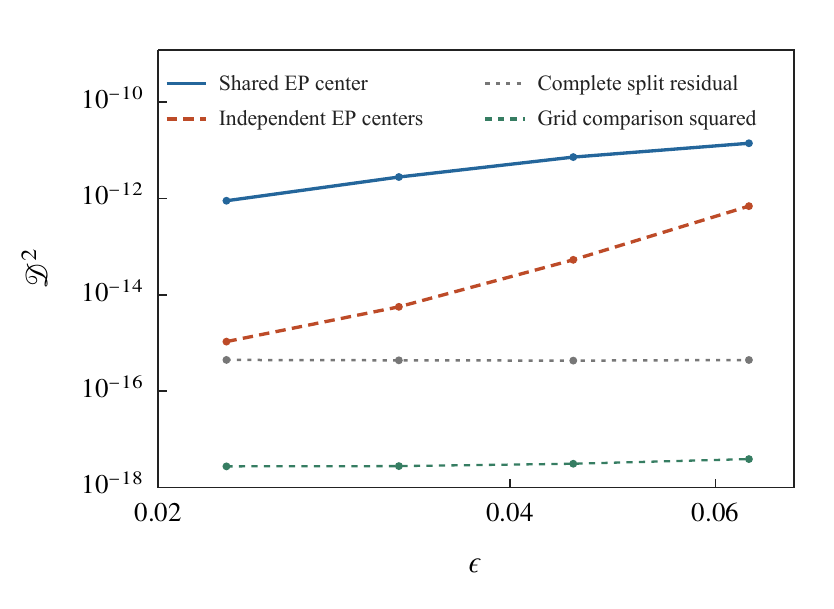}}
    \hfill
    \subfloat[]{\includegraphics[width=.47\textwidth]{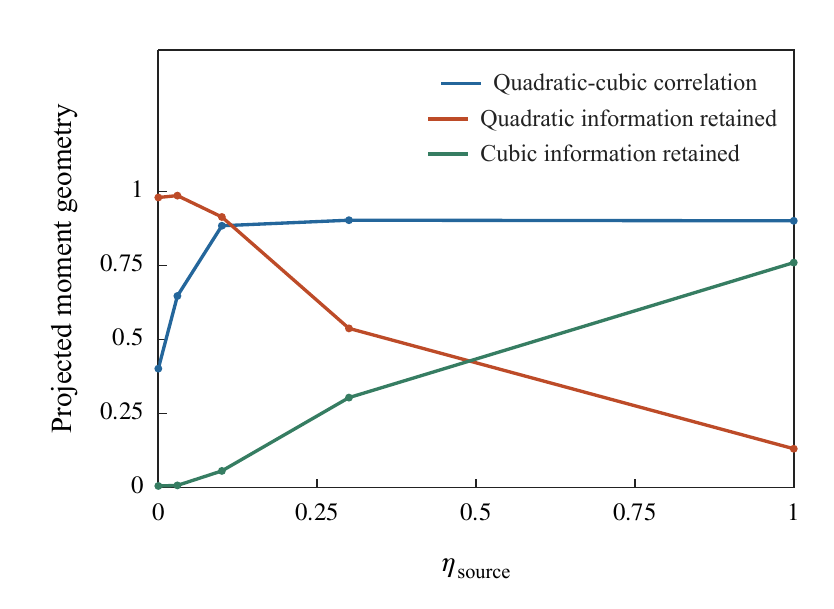}}
    \caption{(a) Normalized squared fit residuals with shared and separately fitted EP centers for the same complete EMS waveforms, $\beta=1$, $\delta_\perp=0.4$, $80\le u/M\le140$. Both fits share ordinary frequencies and use six tail terms. The gray and green curves give the complete split residual and the squared relative difference between successive extrapolated waveforms. (b) Changing the ordinary response through a source direction with zero EP cluster excitation. Curves give the largest normalized quadratic--cubic moment correlation and the information traces retained after profiling ordinary frequencies, relative to keeping those frequencies fixed.}
    \label{fig:s_physical_center}
\end{figure*}

For $\beta=1$, the four recovered separations are $\epsilon=0.02287,0.03214,0.04532,0.06408$. The shared-center $\mathcal D^2$ ranges from $8.95\times10^{-13}$ to $1.40\times10^{-11}$, and the separate-center value from $1.07\times10^{-15}$ to $6.92\times10^{-13}$. At $\epsilon=0.04532$ their ratio is about 136. The complete split residual is about $4.4\times10^{-16}$. The same-source difference between successive Richardson results has relative norm $1.7$--$2.0\times10^{-9}$. The ordinary-response projection varies across the backgrounds, so Fig.~\ref{fig:s_physical_center}(a) shows the distances at finite separation directly.

The role of ordinary responses can be varied independently of the prescribed EP cluster coefficients. Add to the source a direction in the null space of $T$, normalized in the initial-energy metric, and then restore the total energy. The relative change in EP cluster excitation is fixed by this common normalization, while the ordinary response changes. After profiling its shared frequencies, the largest absolute correlation between quadratic and cubic residual directions reaches 0.90 [Fig.~\ref{fig:s_physical_center}(b)]. This is a physical realization of the nonzero cross moments derived below.

\section{Coherent omission and finite-resolution inference}
\label{sec:s_omission}
Figure~2(a) of the Letter uses a unit-energy source $w$ on the split background $q_{\mathrm{BH}}=q_\EP+2\times10^{-4}$, observed at $x_o=60$ over $70\le u/M\le150$. The waveforms are extrapolated from grids $\Delta x=0.025,0.0125$ as in Sec.~VI, and the likelihood follows Sec.~VII. The nominal-EP experiment in Fig.~2(b) instead uses the two energy-constrained sources of Sec.~V.

The Letter's statistical upper bound is $\chi_{\rm up}=\sup_{\ch\in\mathcal C}|\ch|$, with $\mathcal C=\{\ch:\Lambda(\ch)\le t_\chi\}$. Since $\Lambda(0)=T_\chi$, the condition $T_\chi\le t_\chi$ is exactly zero-splitting compatibility. To compute the upper-bound condition without tracing the entire confidence-region boundary, define
\begin{equation}
    T_{\rm res}(r)=\min_{|\ch|\ge r}\Lambda(\ch),\qquad
    r=\chi_*-b_\chi=0.0045.
    \label{eq:s_resolution}
\end{equation}
The finite-resolution EP condition is $T_J>15$, $\Lambda(0)=T_\chi\le10$, and $T_{\rm res}(r)>10$. The exterior minimization varies radius, phase, EP cluster center, and ordinary frequencies together. It uses angular multistarts and radial optimization throughout the specified parameter range of damped responses. This is a finite-resolution criterion for that competing response family.

\subsection{Source and validation for the Letter's omission comparison}
The twelve real coefficients of $w$ specify the Gaussian field and velocity superposition in Sec.~VI and satisfy $w^TEw=1$, where $E$ is the full initial-energy Gram matrix. The full-precision coefficients are supplied in \texttt{omission\_contrast\_source.json}, with source identifier \texttt{t03\_+0.45}. We use $e_j$ for the $j$th coordinate vector, numbered from one. The adaptive search first evaluated 48 deterministic initial-data perturbations. A second stage varied ten coordinates and retuned the first and ninth coefficients to place the omitted-model fit near zero splitting: 40 candidates were followed by 8 extended and 9 local candidates. These are search-stage counts, not draws from a source population. The adopted source was also tested in a separate 50-pair pilot at each of three SNRs, using seed 2026092402. The 200 paired realizations in the Letter use the independent seed 2026092403. This construction demonstrates an attainable omission bias and its local persistence; it does not estimate its occurrence rate in astrophysical sources. At $\rho=2.3\times10^5$, the reference and omitted noise-free fits give $T_\chi=27.53$ and approximately zero, respectively. Their squared residual norms are $0.00303$ and $47.61$.

\par\medskip\noindent
\begin{minipage}{\columnwidth}
    \captionof{table}{Paired fits for source $w$ at $\rho=2.3\times10^5$: the 200-trial ensemble in Fig.~2(a) of the Letter.}
    \label{tab:s_omission_contrast}
    \begin{ruledtabular}
        \begin{tabular}{lrr}
            &$N_{\rm ord}=2$&Omitted fast\\\hline
            Median $T_J$&$4377.48$&$1.19336\times10^6$\\
            Median $T_\chi$&$26.92$&$1.43$\\
            $T_\chi>10$&$196/200$&$2/200$\\
            Full EP criterion&$0/200$&$196/200$\\
            Median split-model RSS&$377.31$&$430.56$\\
        \end{tabular}
    \end{ruledtabular}
\end{minipage}
\par\medskip
All trials have $T_J>15$. Fits include multiple center branches; additional starts from expanded-band solutions and exterior-profile checks near the decision boundary leave the reported classifications unchanged. A conditional residual check uses 400 noise realizations of the best omitted EP waveform for $w$. The 99th percentile of the refitted split-model RSS is $447.48$; 142 of the 200 physical omitted fits fall below it, including 138 that satisfy the false EP criterion. The same EP-null ensemble gives the 99th percentile $T_\chi=9.62$. This conditional residual check measures the overall adequacy of the omitted fit for the source and response family used in Fig.~2(a).

The contrast also persists near $w$. Adding $-0.003e_4/\sqrt{E_{44}}$ or $0.003(e_4/\sqrt{E_{44}}+e_8/\sqrt{E_{88}})$ to $w$ and restoring unit energy gives 49 and 47 omitted-model EP outcomes in separate 50-trial ensembles; the reference models reject zero in 50 and 49 trials. All reference fits give zero EP outcomes. Changing the extrapolation from grids $(0.025,0.0125)$ to $(0.05,0.025)$ changes the recovered splitting by $3.3\times10^{-8}$; the waveform difference has noise-weighted norm $0.0204$ at the adopted SNR.

Figure~2(a) uses the ordinary-frequency bands stated in Sec.~VII. Its best reference EP fit reaches the fast-component boundaries $M\gamma=0.12$ and $M\Omega=0.90$. Enlarging that component's damping and frequency domain to $[0.08,0.50]\times[0.60,1.20]$ reduces the noise-free $T_\chi$ to $2.52$; $[0.02,1.00]\times[0.45,1.80]$ reduces it to $0.783$, with an interior minimum. The slow-component band and all other settings remain fixed. Thus the threshold-crossing comparison is conditional on the specified response family. Enlarging the frequency freedom illustrates the loss of resolution discussed alongside the response-count experiment in the Letter.

\subsection{A local source region for the adopted comparison}
\label{sec:s_current_source_region}
We test a two-dimensional family through the source used in Fig.~2(a), with all waveform, frequency-range, and likelihood settings fixed. Starting from $e_4/\sqrt{E_{44}}$ and $e_8/\sqrt{E_{88}}$, project off $w$ and apply Gram--Schmidt in the energy inner product. The resulting columns of $V$ satisfy $V^TEV=I_2$ and $V^TEw=0$. The family
\begin{equation}
    w(u)=\frac{w+Vu}{\sqrt{1+u^Tu}},\qquad u\in\mathbb R^2,
    \label{eq:s_current_family}
\end{equation}
has unit initial energy. Each whitened waveform is scaled to the specified SNR, as in Fig.~2(a).

The nonzero-area claim follows from strict decision margins and continuity. For any fixed response family $\mathcal M$, including the exterior family $|\ch|\ge r$, the distance $d_{\mathcal M}(y)=\inf_{m\in\mathcal M}\norm{y-m}$ is 1-Lipschitz in $y$. Differences of its squares, including $T_J$, $T_\chi$, and $T_{\rm res}$, are therefore continuous under source perturbations. At $u=0$, the noise-free omitted fit has $T_J=1.194\times10^6$, $T_\chi\simeq0$, and $T_{\rm res}=33.21$, while the reference fit has $T_\chi=27.53$. All four inequalities defining the false-EP contrast are strict. They persist in an open neighborhood of $u=0$, which has nonzero area. This argument concerns the fixed response families; the expanded-frequency comparison in Sec.~IX A uses a different family.

The local area of zero-splitting compatibility can also be obtained analytically. Let $\vartheta=(\Re\ch,\Im\ch)$ denote the best omitted-model splitting, and let $\eta$ collect the nonlinear nuisance parameters. Write $S(\vartheta,\eta;u)$ for the residual sum of squares divided by the squared signal norm, after profiling the linear coefficients. At a regular interior minimum with $\vartheta=0$, define $H=\tfrac12\partial^2_{(\vartheta,\eta)}S$ and
\begin{equation}
    F_0=H_{\vartheta\vartheta}
    -H_{\vartheta\eta}H_{\eta\eta}^{-1}H_{\eta\vartheta},
    \qquad L=\left.\frac{\partial\vartheta}{\partial u}\right|_0.
    \label{eq:s_current_curvature}
\end{equation}
The full Hessian includes residual curvature because the omitted model is misspecified. For a stable minimizing branch, $\vartheta(u)=Lu+O(|u|^2)$, and profiling $\eta$ gives
\begin{equation}
    T_\chi(u)=\rho^2[u^TKu+O(|u|^3)],\qquad K=L^TF_0L.
    \label{eq:s_current_quadratic}
\end{equation}
If $F_0$ is positive definite and $L$ has rank two, the local component with $T_\chi\le t_\chi$ has coordinate area
\begin{equation}
    \mathcal A_{\rm comp}(\rho)
    =\frac{\pi t_\chi}{\rho^2\sqrt{\det K}}[1+o(1)].
    \label{eq:s_current_area}
\end{equation}
The area is measured in the orthonormal source coordinates $du_1du_2$; the intrinsic energy-sphere area has the same leading term. Equation~\eqref{eq:s_current_area} describes zero-splitting compatibility. The full false-EP contrast additionally requires Jordan evidence, a tight upper bound, and reference-model rejection.

For the adopted source, the best omitted splitting is zero to numerical precision. The energy Gram matrix, tangent vectors, and full-precision derivatives are supplied with the data. Numerically,
\begin{equation}
    \begin{aligned}
        L&\simeq\begin{pmatrix}0.057714&0.022693\\-0.003490&0.008710\end{pmatrix},\\
        F_0&\simeq10^{-5}\begin{pmatrix}3.81630&0.44198\\0.44198&3.44301\end{pmatrix}.
    \end{aligned}
\end{equation}
Both conditions in Eq.~\eqref{eq:s_current_area} hold. Central source differences reproduce $L$ to relative error below $7\times10^{-5}$, and halving the Hessian step changes $H$ by less than $2\times10^{-8}$. The predicted area coefficient is $\pi t_\chi/\sqrt{\det K}\simeq1.501\times10^9$. At $\rho_0=2.3\times10^5$, the quadratic ellipse has semiaxes $0.2516$ and $0.03589$ in $u$.

We test the actual profiles on a $9\times9$ grid in coordinates $z$, where $u=U\operatorname{diag}(a_1,a_2)z$ and $U,a_i$ are the ellipse axes at $\rho_0$. Each coordinate runs from $-1.2$ to $1.2$ in steps of $0.3$. Of 81 points, 37 permit zero splitting in the omitted fit; 28 also satisfy its full EP criterion and reject zero in the reference fit. All 13 grid points with $|z|\le0.6$ show this contrast. Over those points, omitted-model $T_\chi\le4.13$, $T_{\rm res}\ge10.84$, and $T_J\ge1.008\times10^6$, while reference-model $T_\chi\ge20.93$. The grid checks finite source perturbations; the continuity argument supplies the open-neighborhood conclusion between sampled points. These deterministic grid counts are not source-population frequencies.

\begin{figure*}[t]
 \centering
 \subfloat[]{\includegraphics[width=.47\textwidth]{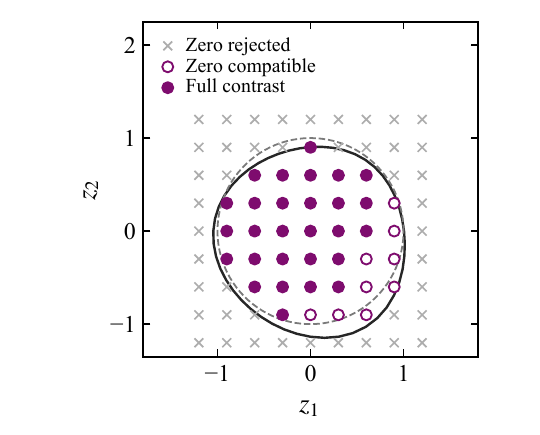}}
 \hfill
 \subfloat[]{\includegraphics[width=.47\textwidth]{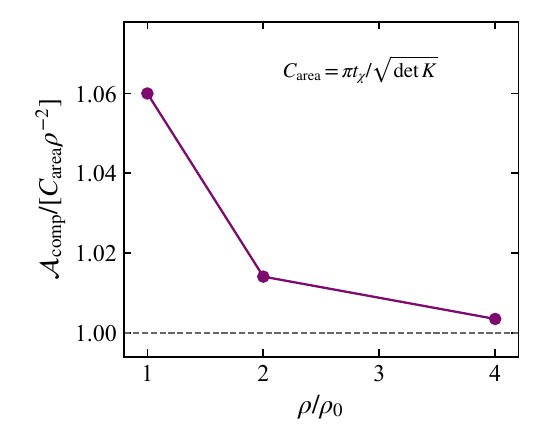}}
 \caption{Local source geometry for the source in Fig.~2(a). (a) Noise-free fits at $\rho_0=2.3\times10^5$ in normalized ellipse coordinates. Filled circles satisfy the omitted-model full EP criterion and reference-model rejection of zero; open circles permit zero in the omitted fit but fail another requirement of this contrast. Crosses reject zero in the omitted fit. The solid boundary is the nonlinear $T_\chi=10$ profile; the dashed circle is its quadratic approximation. (b) Numerically integrated zero-compatible area divided by the $\rho^{-2}$ prediction. Its approach to unity verifies the local area law. Full EP identification requires the additional cuts illustrated in panel (a).}
 \label{fig:s_current_source_region}
\end{figure*}

To test the area law, we solve $T_\chi=10$ along 48 angular rays, using full nonlinear fits. At $\rho/\rho_0=1,2,4$, the areas are $0.030071$, $0.00719199$, and $0.00177917$, respectively. Their ratios to Eq.~\eqref{eq:s_current_area} are $1.0600$, $1.0141$, and $1.0035$. Doubling the angular sampling from 24 to 48 changes each area by less than $3\times10^{-8}$ in relative terms; this is an angular-integration check, not an estimate of all numerical errors. Figure~\ref{fig:s_current_source_region} shows the finite-SNR region and convergence to the asymptotic law.

\subsection{Nominal-EP calibration and recovery controls}
The likelihood thresholds are calibrated for the free-center fits, whose simple-pole limit is singular \cite{Drton2009}. For the nominal-EP two-source configuration of Sec.~V, 1500 realizations of the single-exponential model at $\rho=10^5$ give a 99th percentile of $T_J=13.22$; the adopted threshold is 15. On eight complex directions at $|\ch|=0.005$, 800 realizations at $\rho=150000$ give a 99th percentile of $\Lambda(\ch_{\rm true})=8.34$ and coverage $798/800$ at threshold 10. For the unit-energy single source selected in Sec.~V at $\rho=3\times10^6$, the corresponding calibrations for the single-exponential model and the splitting boundary use 500 and 400 realizations; their 99th percentiles are $8.37$ and $8.85$. These calibrations specify the operating point for the tested configurations.

\par\medskip\noindent
\begin{minipage}{\columnwidth}
 \captionof{table}{Finite-resolution identification of the nominal EP. Intervals are 95\% Wilson intervals for the measured recovery fraction. The target is $\chi_*=0.005$ with $b_\chi=0.0005$.}
 \label{tab:s_power}
 \setlength{\tabcolsep}{1.5pt}
 \begin{ruledtabular}\begin{tabular}{lrrl}
 Source (Sec.~V)&$\rho$&Count&Fraction [interval]\\\hline
 Two sources&$3\times10^5$&$147/200$&$.735\ [.670,.791]$\\
 & $4\times10^5$&$191/200$&$.955\ [.917,.976]$\\
 & $5\times10^5$&$199/200$&$.995\ [.972,.999]$\\
 Selected single&$5.288\times10^5$&$0/200$&$0\ [0,.0188]$\\
 &$1.2\times10^6$&$15/200$&$.075\ [.046,.120]$\\
 &$3\times10^6$&$195/200$&$.975\ [.943,.989]$\\
 Single Gaussian&$2\times10^6$&$79/200$&$.395\ [.330,.464]$\\
 &$4\times10^6$&$197/200$&$.985\ [.957,.995]$\\
 \end{tabular}\end{ruledtabular}
\end{minipage}
\par\medskip
Table~\ref{tab:s_power} reports a separate SNR-scan ensemble of noisy waveforms from the wave equation. The two-source value $\rho=4\times10^5$ and selected-single value $528778.935$ correspond to the same initial energy and noise amplitude through Eq.~\eqref{eq:s_distances}. Their recovery fractions are $191/200$ and $0/200$. The response-count comparison below uses its own paired draws and gives $190/200$ at the same two-source operating point. At $\rho=5\times10^5$, both $q_{\mathrm{BH}}-q_\EP=\pm2\times10^{-4}$ give resolved splitting in all 200 trials and no finite-resolution EP identifications.

Readouts $x_o=30,60$, window starts $u_0/M=70,75,80$, and four or five tail terms give maximum nominal-EP displacements $|\ch|=1.27\times10^{-4}$ for the two-source configuration and $1.54\times10^{-4}$ for the selected single source. The allowance $b_\chi=5\times10^{-4}$ covers this empirical variation. With a fifth tail term, the two-source noise-free threshold crossing rises from $2.59\times10^5$ to $2.77\times10^5$; actual recovery at $\rho=5\times10^5$ is $200/200$. A 400-realization boundary calibration for that model gives 99th percentile $7.25$ and coverage $400/400$ at threshold 10. The quoted allowance refers to these waveform and numerical controls.

The noise-free difference from the single-exponential model crosses $T_J=15$ near $\rho=8406$. This distance forecast and the actual finite-resolution recovery rates measure different information: the first detects Jordan time dependence, while the recovery criterion constrains the complex splitting after all ordinary responses are fitted.

\subsection{Changing the number of fitted ordinary response terms}
Figure~2(b) of the Letter varies the number $N_{\rm ord}$ of fitted ordinary response terms from zero to five for the nominal-EP two-source configuration of Sec.~V. The source coefficients are supplied in \texttt{nominal\_ep\_pair\_source.json}. The experiment uses $x_o=60$, $70\le u/M\le150$, and four tail terms. The first two ordinary frequency bands are those of Sec.~VII. Response terms three through five have $(M\gamma,M\Omega)$ ranges $[0.20,0.36]\times[0.74,0.84]$, $[0.28,0.44]\times[0.81,0.90]$, and $[0.36,0.54]\times[0.88,0.98]$. The additional frequencies and all amplitudes are fitted. The count excludes the EP cluster and tails. All six models use the same 200 noise realizations at $\rho=4\times10^5$, with seed 2026092217. Table~\ref{tab:s_mode_count} lists the full-criterion outcomes. Fits with additional ordinary response terms frequently reach ordinary-frequency boundaries, so these outcomes include the stated frequency-range assumptions.

\par\medskip\noindent
\begin{minipage}{\columnwidth}
    \captionof{table}{Finite-resolution EP outcomes for the nominal-EP two-source signal in Fig.~2(b) of the Letter. The 200 paired trials use $\rho=4\times10^5$, $(t_J,t_\chi)=(15,10)$, $\chi_*=0.005$, and $b_\chi=0.0005$.}
    \label{tab:s_mode_count}
    \begin{ruledtabular}
    \begin{tabular}{cr}
    $N_{\rm ord}$ & EP outcomes\\
    0 & $0/200$\\
    1 & $15/200$\\
    2 & $190/200$\\
    3 & $1/200$\\
    4 & $1/200$\\
    5 & $0/200$
    \end{tabular}
    \end{ruledtabular}
\end{minipage}
\par\medskip
For the nominal-EP noise-free waveform at $\rho=4\times10^5$, the improvement in best split residual square is 214.1 when the second ordinary response term is added, and 0.049 when the third is added. Yet the exterior statistic at radius 0.0045 falls from 23.84 for two terms to 1.636 for three and 0.166 for five. This is the loss of exclusion power described in the Letter. If several physically allowed models remain viable, a conservative splitting region is the union of their profile regions. Coverage for the true member is retained when that member remains in the candidate set; any data-dependent exclusion rule also needs calibration.

\subsection{Signal with three ordinary response terms}
The tensor injection uses the signal, noise, and antenna conventions of Sec.~XI, with $f_c=150$ Hz, $Q=3$, $\epsilon=0.16$, $\delta_\perp=0$, and $\rho=10^4$. Three ordinary terms have dimensionless frequencies $-2.1+7.4\ii$, $-3+8.2\ii$, and $-4+9\ii$, and share polarization direction $c$. Their amplitudes are $A_o$, $0.015e^{0.4\ii}ae^{\ii\varphi}$, and $0.01e^{-0.8\ii}ae^{\ii\varphi}$, where $A_o$ is specified in Eq.~\eqref{eq:s_strain_ordinary}, $a=0.3,1,3$, and $\varphi=0,\pi/2,\pi$. Each retained frequency varies over the damping and oscillation bands stated in Sec.~XI. Table~\ref{tab:s_three_subsets} compares all eight subsets across these nine signals. The full response recovers the true $\ch=-0.0256$ with maximum absolute error $3.2\times10^{-12}$. Large incomplete-model estimates can reach parameter boundaries; their range diagnoses model dependence.

\par\medskip\noindent
\begin{minipage}{\columnwidth}
    \captionof{table}{Noise-free fitted splitting magnitude over nine amplitude--phase settings, with true $|\ch|=0.0256$. The retained subset identifies the ordinary components by their order above.}
    \label{tab:s_three_subsets}
    \begin{ruledtabular}
    \begin{tabular}{crr}
    Retained subset & Minimum $|\ch|$ & Maximum $|\ch|$\\
    $\varnothing$ & 0.009193 & 0.353553\\
    $\{1\}$ & 0.003782 & 0.112384\\
    $\{2\}$ & 0.017095 & 0.025994\\
    $\{3\}$ & 0.017658 & 0.024595\\
    $\{1,2\}$ & 0.025341 & 0.025659\\
    $\{1,3\}$ & 0.025463 & 0.025611\\
    $\{2,3\}$ & 0.025156 & 0.025454\\
    $\{1,2,3\}$ & 0.025600 & 0.025600
    \end{tabular}
    \end{ruledtabular}
\end{minipage}
\par\medskip

A complementary coherent example fixes the second and third amplitudes at $0.015e^{0.4\ii}$ and $0.01e^{-0.8\ii}$, and sets the first to approximately $-0.005824-0.011136\ii$. At $\rho=2\times10^4$, retaining only component 1 satisfies the false finite-resolution criterion in $112/200$ paired trials, using radius 0.02 and $(t_J,t_\chi)=(15,10)$. Retaining components 1 and 2 or all three gives zero such outcomes. Their split-detection counts are respectively $178/200$ and $41/200$, illustrating the precision cost of additional fitted parameters. Absolute residuals supply a separate model check: omitting all three gives mean split residual square 1376 for 482 data values, whereas retaining component 1 gives 531. These examples quantify the stated decision rule; a combined model-selection and EP-identification error rate requires calibration of both operations.

\subsection{Conditional threshold checks for additional models}
For each of $N_{\rm ord}=2,3,5$, calibration signals are generated from an exact fit with two ordinary response terms, with the nominal-EP source amplitudes, $\rho=4\times10^5$, and the noise of Sec.~VII. Additional fitted components have zero true amplitudes. Training uses 1000 trials of the single-exponential model, 1000 EP trials, and 250 trials on each of eight complex splitting directions at $|\ch|=0.005$. Independent validation uses 500 trials each of the single-exponential and EP models, plus 125 on each boundary direction. Table~\ref{tab:s_conditional_calibration} gives exceedances at the declared thresholds 15 and 10, which remain fixed in the response-count experiment. The boundary statistic is the profile loss at the true splitting. These checks are conditional on the specified signals and parameter ranges, and do not establish uniform coverage over all source amplitudes, frequencies, or model-selection procedures.

\par\medskip\noindent
\begin{minipage}{\columnwidth}
    \captionof{table}{Independent validation at fixed thresholds. Columns give false Jordan detection for a single-exponential model, exclusion of zero for an EP, and exclusion of the true boundary splitting.}
    \label{tab:s_conditional_calibration}
    \begin{ruledtabular}
    \begin{tabular}{crrr}
    $N_{\rm ord}$ & $T_J>15$ & $T_\chi>10$ & $\Lambda(\ch_{\rm true})>10$\\
    2 & 5/500 & 3/500 & 1/1000\\
    3 & 0/500 & 3/500 & 3/1000\\
    5 & 0/500 & 1/500 & 0/1000
    \end{tabular}
    \end{ruledtabular}
\end{minipage}
\par\medskip

\section{Ordinary-response overlap and spectral information}
After existing nuisance parameters are removed in the noise inner product, let $U$ be the subspace spanned by the waveform derivatives with respect to splitting and $V$ the subspace spanned by derivatives with respect to ordinary-response parameters. The smallest principal angle $\theta$ measures the closest alignment of directions in these two subspaces. Projection geometry gives \cite{ScharfFriedlander1994}
\begin{equation}
    \begin{aligned}
        \eta_{\min}&=\inf_{u\in U\setminus0}
        \frac{\norm{(I-P_V)u}^2}{\norm{u}^2}=\sin^2\theta,\\
        \mathcal K&=\sup_{b\in V\setminus0}
        \frac{\norm{P_Ub}}{\norm{(I-P_U)b}}=\cot\theta,\\
        \eta_{\min}&=\frac1{1+\mathcal K^2}.
    \end{aligned}
    \label{eq:s_angle}
\end{equation}
Thus an omitted component that is efficiently absorbed along the splitting direction also removes substantial splitting information when fitted. The local displacement follows the usual waveform-error projection \cite{CutlerVallisneri2007}; in the EP problem the same coherent response can also project onto the Jordan direction.

For complex data with white noise at positive times, the determinant of the exponential Gram matrix and its frequency derivatives makes this cost explicit. Projecting $g_n=\tau^ne^{s\tau}$ off $g_0,\ldots,g_{n-1}$ and the functions $\tau^r e^{\nu_k\tau}$, $0\le r<m_k$, at each ordinary pole $\nu_k$ yields
\begin{equation}
    \norm{Qg_n}^2=\frac{(n!)^2}{(-2\Ree s)^{2n+1}}
    \prod_k\left|\frac{s-\nu_k}{s+\bar\nu_k}\right|^{2m_k}.
\end{equation}
Here $m_k$ counts the functions retained at that pole. For a single record, $m_k=1$ removes its complex amplitude, while $m_k=2$ also removes the nonzero-residue frequency tangent. These rational factors also occur in QNM filters \cite{MaFilters2022,MaCleaning2023}. Shared frequencies across records instead require the coupled tangent projection in Eq.~\eqref{eq:s_weighted}.

\subsection{Real moment geometry after nuisance projection}
For a real measured response, apply the radiation map $\mathcal R$, whiten with $W_N$, and project out the fitted amplitude and ordinary-response tangents. Near a matching center branch, the next center adjustment also removes the quadratic moment in the direction $c$. Denote the resulting projection by $Q$. The four real residual columns are
\begin{equation}
    M=QW_N\mathcal R\,[vg_2,\ii vg_2,cg_3,\ii cg_3],
    \qquad G=M^TM.
    \label{eq:s_real_gram}
\end{equation}
The notation $\mathcal R$ includes taking the real detector response. With $\alpha=-a_0\epsilon^2\delta_\perp\norm{c}$ and $\beta=a_0\epsilon^3\Delta/3$, Eq.~\eqref{eq:s_two_residuals} gives
\begin{equation}
    d_N^2\sim z^TGz,\qquad
    z=(\Ree\alpha,\Imm\alpha,\Ree\beta,\Imm\beta)^T,
    \label{eq:s_real_quadratic}
\end{equation}
followed by minimization over the competing center branches. The Gram matrix determines which residual structures survive the actual readout. Its off-diagonal blocks can contribute at order $\epsilon^5\delta_\perp$. The ideal $1+\Xi^2$ curve consequently becomes a quadratic form in the retained moments.

An exact white-noise example shows how a shared ordinary frequency produces this coupling. Take $c=(1,0)$, $v=(0,1)$, and an ordinary pole $-2$. Remove $g_0,g_1,g_2,e^{-2x}$ from the first record and $g_0,g_1,e^{-2x}$ from the second. The ordinary amplitudes are independent, but the frequency tangent is $(xe^{-2x},xe^{-2x})$. Projecting it out gives the Gram matrix for the remaining quadratic and cubic directions,
\begin{equation}
    G=\begin{pmatrix}1/360&1/180\\1/180&41/1440\end{pmatrix},
    \qquad \det G=\frac1{20736}.
    \label{eq:s_exact_cross_gram}
\end{equation}
The nonzero off-diagonal element persists even though $c$ and $v$ were orthogonal before ordinary-frequency profiling. Figure~\ref{fig:s_physical_center}(b) shows the corresponding effect for complete EMS waveforms.

\subsection{Bias from an omitted ordinary component}
After removing the existing nuisance tangents, let $D_\chi$ contain the two real splitting derivatives and let $b_{\mathrm{miss}}$ be an omitted coherent response. The local parameter bias and the information after including its tangent space $V$ are \cite{CutlerVallisneri2007}
\begin{equation}
    \begin{aligned}
        \Delta\vartheta&\simeq(D_\chi^TD_\chi)^{-1}D_\chi^Tb_{\mathrm{miss}},\\
        F_{\mathrm{full}}&=D_\chi^T(I-P_V)D_\chi.
    \end{aligned}
    \label{eq:s_bias_information}
\end{equation}
The projection $D_\chi^Tb_{\mathrm{miss}}$ fixes the bias direction. Equation~\eqref{eq:s_angle} describes the largest possible bias-to-residual ratio allowed by the ordinary subspace and its corresponding information loss. In the physical source plane, the minimum retained fraction of information is $0.00723$--$0.00795$; the displacement calculated from the actual omitted waveform follows the nonlinear bias with median relative difference 0.126. The full nonlinear profiles determine the source boundaries and likelihood statistics reported above.

\section{Tensor strain and detector-noise inference}
The detector experiment evaluates the projected EP cluster response directly in the two real gravitational-wave polarizations. In dimensionless time, the injected polarization vector is
\begin{equation}
    H(x)=e^{(-1+6\ii)x}\left[c\cos(\epsilon x)+k\sin(\epsilon x)\right],
    \label{eq:s_tensor_target}
\end{equation}
where $c=(1,1)/\sqrt2$ and $k=(\lambda+\delta_\perp,\lambda-\delta_\perp)/\sqrt2$, with $\lambda=1/2$. Thus $j=\epsilon k$, $\ch=-\epsilon^2$, and $Q=\Omega_c/(2\gamma_c)=3$. Independent degenerate radiation channels realize this excitation family as described above. The measured data are $(\Ree H)F^T$. We use ideal co-located right-angle interferometers with
\begin{equation}
    F=\begin{pmatrix}1&0\\\cos2\alpha&\sin2\alpha\end{pmatrix},
    \qquad \alpha=0^\circ,15^\circ,45^\circ.
    \label{eq:s_detector_antenna}
\end{equation}
This geometry controls the retained polarization rank. Polarization coefficients are shared through $F$ in every fitted model.

\subsection{Finite-window covariance and profile likelihood}
The public aLIGO design sensitivity file \cite{LIGODesignT1800044} supplies an amplitude spectral density $A(f)$. We set the one-sided noise power spectral density to $S_n(f)=A(f)^2$, interpolate its logarithm, and hold the value at 5 Hz constant below the tabulated lower endpoint. We sample $x=0,0.05,\ldots,12$, giving 241 time samples, with $\Delta t=0.05/\gamma_c$ and $\gamma_c=2\pi f_c/6$. For each $f_c=40,80,150,300$ Hz, the discrete autocovariance is calculated from the spectrum up to the sampling Nyquist frequency. With $f_s=1/\Delta t$ and transform size $n_f=65536$,
\begin{equation}
    K_{ab}=\frac{f_s}{2}\,\mathrm{IFFT}_{\mathrm{real}}
    \left[\frac{S_n(f)}{S_{\mathrm{ref}}}\right]_{|a-b|}.
    \label{eq:s_psd_covariance}
\end{equation}
Here $S_{\mathrm{ref}}$ is the minimum tabulated PSD, absorbed into the overall noise scale. The finite Toeplitz matrix $K$ is factored as $K=LL^T$. Whitening uses $L^{-1}$ on the retained samples, so the likelihood contains the inverse of the finite-window covariance \cite{IsiFarr2021}. Detector noises are independent and have the same spectrum in this comparison.

For nonlinear parameters $\theta$, a whitened design matrix $B(\theta)$ contains the real and imaginary quadratures of both physical polarizations. Linear amplitudes are profiled by its scaled singular-value decomposition. The residual is $r=(I-BB^+)d$, and its derivative is
\begin{equation}
    \partial_i r=-(I-BB^+)(\partial_iB)\widehat a
    -(B^+)^T(\partial_iB)^Tr,
    \label{eq:s_variable_projection}
\end{equation}
with $\widehat a=B^+d$. We retain the residual-dependent term. The EP cluster functions and derivatives use their globally convergent power series near $\chi=0$, with real quadratures ordered as $(\Ree f,-\Imm f)$. Fitted dimensionless damping lies in $[0.5,1.6]$, frequency in $[5,7]$, and each component of $\ch$ in $[-0.25,0.25]$. The fit parameterization uses $\chi=\gamma_{\mathrm{fit}}^2\ch$; the target time and its quoted separation use $\gamma_c$. This is a reparameterization of the same response family.

The center minimization uses five starting points: the seed and shifts $\pm0.15$ in log damping or frequency. The split fit additionally starts at the best nested EP solution. In fits that include an ordinary response, its damping range is 0.7--1.4 times its injected value and its frequency varies by $\pm0.35$. Frequencies are shared across detector outputs.

The distance scan contains 624 points: four central frequencies, three antenna geometries, $\delta_\perp=0,0.02,0.1,0.4$, and 13 logarithmically spaced separations from 0.02 to 0.3. If $E_h=\norm{h}_K^2$, the free-center threshold at distance noncentrality $\Lambda_*$ is
\begin{equation}
    \rho_{\mathrm{req}}=\sqrt{\Lambda_*E_h/d_{\EP}^2},
    \qquad \rho^2=E_h/\sigma_w^2.
\end{equation}
For center information we minimize the joint likelihood penalty, including $|q-q_{\mathrm{obs}}|^2/\sigma_q^2$, and solve for the SNR at which that minimum reaches $\Lambda_*$. The curves use $\Lambda_*=10$; Jordan-detection curves use the noise-free squared-residual difference between the single-exponential and EP models, set to 15. Figure~\ref{fig:s_detector_noise}(a) shows how the antenna rank affects the same noncollinear target at fixed total SNR. Increasing Fourier resolution changes representative distances by about 1\% or less; halving the time step, with the associated larger Nyquist frequency, changes them by a few percent.

\subsection{Noisy injections}
At $f_c=150$ Hz and $\epsilon=0.16$, the two frequencies differ by 8 Hz and the damping time is $1/\gamma_c=6.366$ ms. The three distance thresholds $\Lambda_*=10$ are $\rho_0=4419.89$ for collinear excitation with a free center, $1102.32$ for $\delta_\perp=0.4$, and $500.83$ for collinear excitation with $\sigma_q=0.01$. At each of $\rho/\rho_0=1,1.5,2$, we generate 400 EP-null and split injections with paired independent Gaussian draws in whitened coordinates. The EP-null has the same $c,j$ and $\chi=0$. When present, $q_{\mathrm{obs}}=q_c+\sigma_q(z_1+\ii z_2)$ is also randomized with independent standard-normal $z_1,z_2$.

\begin{figure*}[t]
    \centering
    \subfloat[]{\includegraphics[width=.47\textwidth]{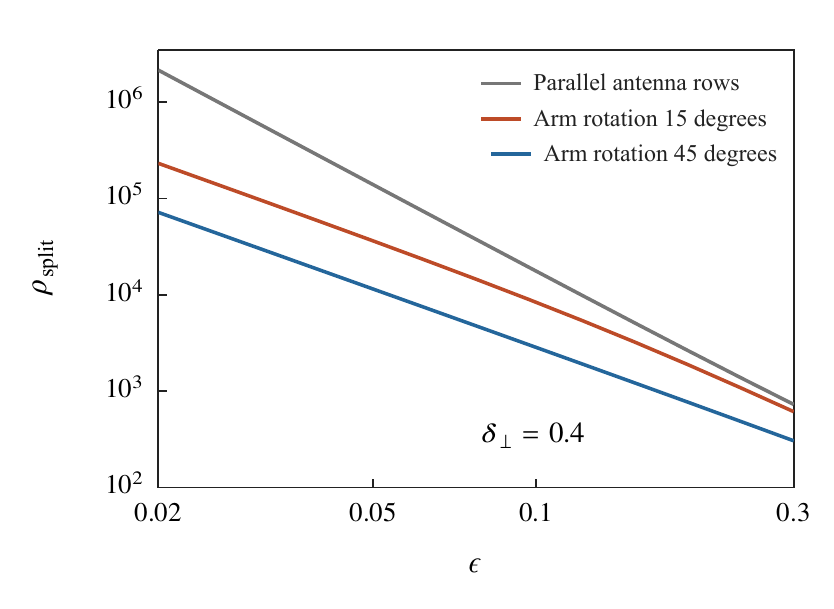}}
    \hfill
    \subfloat[]{\includegraphics[width=.47\textwidth]{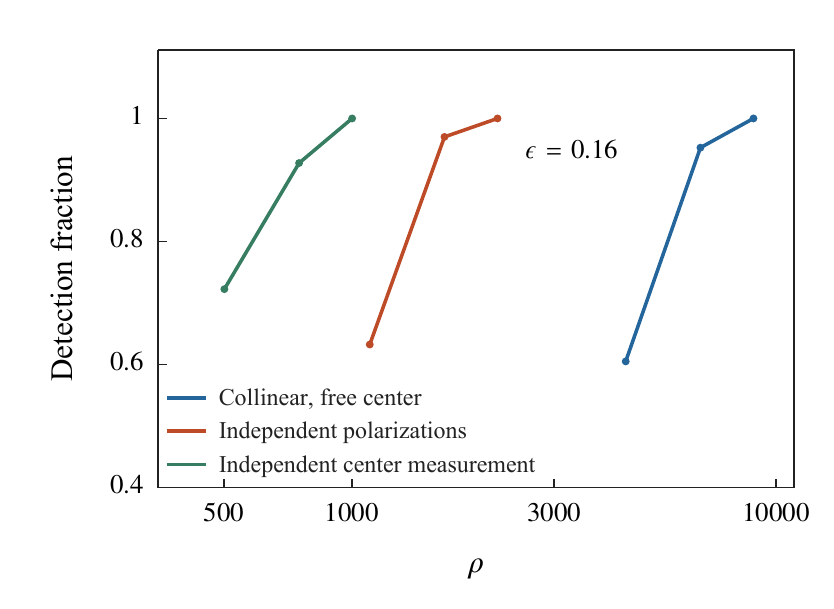}}
    \caption{(a) Strain SNR needed for distance noncentrality 10 with $\delta_\perp=0.4$, $f_c=150$ Hz and the antenna matrix in Eq.~\eqref{eq:s_detector_antenna}. Parallel antenna rows retain only one real polarization direction. (b) Split-signal recovery at $\epsilon=0.16$, using the respective empirical 99\% EP-null thresholds. Each point has 400 split injections. The curve for independent polarizations uses distinct relative excitation compositions with $\delta_\perp=0.4$. The external center measurement is randomized. Lines connect sampled SNRs.}
    \label{fig:s_detector_noise}
\end{figure*}

The empirical 99th percentile of the 400 EP-null values of $T_\chi$ supplies the threshold at each operating point. Table~\ref{tab:s_detector_power} gives the actual split-injection counts, and Fig.~\ref{fig:s_detector_noise}(b) shows their detection fractions. Binomial Wilson intervals supplied with the data are conditional on the estimated threshold. The distance noncentrality describes the noise-free signal separation; the injection table gives the finite-noise detection probability for this calibration.

\begin{table}[htbp]
    \caption{Tensor split detection at $\epsilon=0.16$. Free means no independent center measurement. Each row has 400 EP-null and 400 split injections.}
    \label{tab:s_detector_power}
    \begin{ruledtabular}
    \begin{tabular}{ccrrr}
        $\delta_\perp$ & $\sigma_q$ & $\rho$ & Null 99\% & Detected\\
        0 & Free & 4420 & 9.666 & 242/400\\
        0 & Free & 6630 & 10.528 & 381/400\\
        0 & Free & 8840 & 8.546 & 400/400\\
        0.4 & Free & 1102 & 8.317 & 253/400\\
        0.4 & Free & 1653 & 8.215 & 388/400\\
        0.4 & Free & 2205 & 9.341 & 400/400\\
        0 & 0.01 & 501 & 8.481 & 289/400\\
        0 & 0.01 & 751 & 10.730 & 371/400\\
        0 & 0.01 & 1002 & 9.823 & 400/400
    \end{tabular}
    \end{ruledtabular}
\end{table}

\subsection{Coherent ordinary response in strain}
To the collinear target in Eq.~\eqref{eq:s_tensor_target} with $\epsilon=0.16$, add
\begin{equation}
    \begin{aligned}
        H_{\mathrm{ord}}(x)&=A_o e^{(-2.1+7.4\ii)x}c,\\
        A_o&=-0.00808879-0.01128994\ii.
    \end{aligned}
    \label{eq:s_strain_ordinary}
\end{equation}
The amplitude is chosen so that the noise-free omitted-model best-fit splitting is near zero. The ordinary component has amplitude magnitude $|A_o|=0.0139$ relative to the EP cluster's constant coefficient at the reference time. All components pass through the same antenna and noise response. The complete model fits the ordinary amplitude and complex frequency, and the omitted model lacks that component. Scaling $A_o$ from 0.75 to 1.25 times its baseline value gives the curves in the Letter.

At the baseline amplitude and $\rho=10^4$, the omitted model has $T_J=8292.8$ and $T_\chi\simeq0$. Minimizing the split profile over $|\ch|\ge0.02$ gives $T_{\mathrm{res}}=12.068$, so it satisfies the finite-resolution condition with $(t_J,t_\chi)=(15,10)$ and $b_\chi=0$. The complete model gives $T_J=599.4$ and $T_\chi=12.289$, with its best fit at the true $\ch=-0.0256$. The zero allowance here corresponds to an exactly specified injected response. Exterior profiles at radii $0.008,0.012,0.02$ vary radius, phase and all permitted frequencies, extending to the circumscribed radius $0.25\sqrt{2}$ of the split-parameter domain. These noise-free likelihood penalties demonstrate the bias mechanism in tensor strain; their threshold values are the specified decision thresholds.

\section{Axial gravitational propagation and radiative extraction}
We consider the odd-parity $\ell=2$ Schwarzschild master equation with the localized propagation correction studied in Ref.~\cite{YangBertiFranchini2025}. Set $M=1$, $x=r+2\ln(r-2)$, and
\begin{equation}
    \begin{aligned}
        (\partial_t^2-\partial_x^2+V)\Psi&=0,\\
        V(x)&=\left(1-\frac2r\right)
        \left(\frac6{r^2}-\frac6{r^3}\right)+a_b e^{-(x-d)^2}.
    \end{aligned}
    \label{eq:s_rw_potential}
\end{equation}
The master variable is in the radiative CPM normalization, with the relation to odd-parity strain in Eq.~\eqref{eq:s_cpm_strain}. The Gaussian term specifies an effective axial propagation model; excitation is specified by finite-energy initial data.

For time dependence $e^{-\ii\omega t}$, let $z=\Psi'/\Psi$ and $w=\partial_\omega z$. The Jost solutions, selected by the radiative boundary condition at each end, obey
\begin{equation}
    z'=V-\omega^2-z^2,\qquad w'=-2\omega-2zw.
    \label{eq:s_jost_riccati}
\end{equation}
We integrate inward from exterior complex contours with asymptotic data $(z,w)=(\pm\ii\omega,\pm\ii)$ on the right and left. At an interior point, $F=z_L-z_R$ is the matching function and $F_\omega=w_L-w_R$. The double-zero conditions $F=F_\omega=0$ give
\begin{equation}
    \begin{aligned}
        \omega_*&=0.3653687227-0.1174405994\ii,\\
        a_b&=0.005082941393,\qquad d=15.69758666.
    \end{aligned}
    \label{eq:s_rw_ep}
\end{equation}
Here $F_{\omega\omega}\ne0$. The contours begin at real locations $-35$ and 30 and turn by 0.6 radians; their exterior lengths are 60 and 120. The right-hand inversion of the tortoise coordinate is continued along the contour by solving $r+2\ln(r-2)=x$ on the same branch. Integration to the matching point $x=3$ supplies both frequency derivatives. Contour angles 0.5--0.7 and right lengths 120--160 give matching residual below $10^{-7}$ and derivative residual below $10^{-5}$.

With $a_b=0$, the fundamental frequency is $0.37367168445-0.08896231569\ii$. Increasing $a_b$ from its resonant value by 10\% gives the two poles
\begin{equation}
    \begin{aligned}
        \omega_+&=0.3779843340-0.1191450445\ii,\\
        \omega_-&=0.3534821956-0.1123176152\ii.
    \end{aligned}
\end{equation}
The positive-phase waveform convention is $s=(-\ii\omega)^*$. The independently determined EP cluster center is therefore the mean of these two poles after that conversion.

\begin{figure*}[t]
    \centering
    \subfloat[]{\includegraphics[width=.47\textwidth]{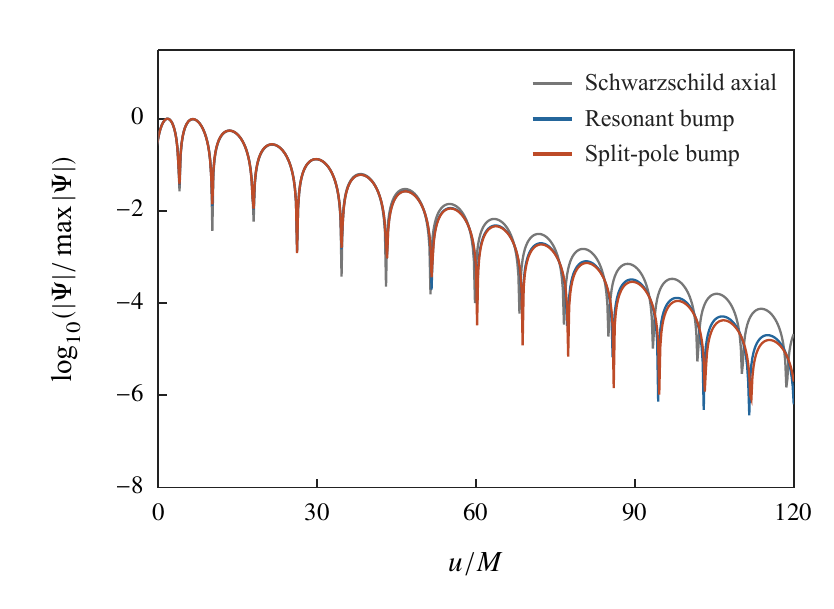}}
    \hfill
    \subfloat[]{\includegraphics[width=.47\textwidth]{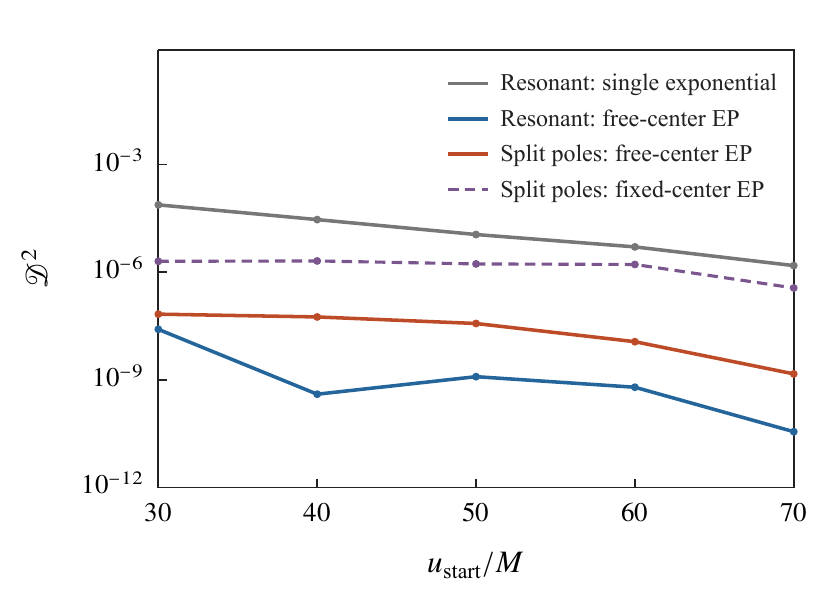}}
    \caption{(a) Radiative CPM waveforms from a unit-energy Gaussian displacement in the Schwarzschild potential, the resonant Gaussian-corrected potential, and the potential with a split pair obtained by increasing $a_b$ by 10\%. Each curve is normalized to its peak before taking the absolute value and logarithm. (b) Normalized squared fit distances versus window start, with fixed end $u=110$. All models retain the same ordinary response and three tail functions. The split-potential fixed center is the mean of its independently computed Jost poles.}
    \label{fig:s_gravitational}
\end{figure*}

For each potential, four Gaussian displacement or velocity profiles at $(p,w)=(0,2),(10,2)$ are normalized by the initial-energy form in Eq.~\eqref{eq:s_energy}. They are evolved on $[-600,600]$ with $\Delta x=0.1,0.05,0.025$, $\Delta t=\Delta x/2$, up to $t=330$. Values at $x_o=100,200$ are aligned at common retarded time $u=t-x_o$. If $r_1,r_2$ are the corresponding areal radii, the leading radiative master amplitude is extracted as
\begin{equation}
    \Psi_{\infty}(u)=
    \frac{r_2\Psi(u,r_2)-r_1\Psi(u,r_1)}{r_2-r_1}
    +O((r_1r_2)^{-1}).
    \label{eq:s_radial_extrapolation}
\end{equation}
This removes the first $1/r$ correction to the master variable. The strain retains the explicit overall $1/r$ in Eq.~\eqref{eq:s_cpm_strain}. Figure~\ref{fig:s_gravitational}(a) shows $\log_{10}|\Psi_\infty|$ normalized to each waveform's peak. All fits use the signed radiative field.

The two-source fits use the displacement at $p=0$ and the velocity at $p=10$, each with half of the total initial energy. The model contains the EP cluster, one ordinary response term with free shared frequency, and three tail functions of the same finite-window form as Eq.~\eqref{eq:s_model}. The ordinary frequency bands are $\gamma\in[0.13,0.4]$ and $\Omega\in[0.53,0.85]$. EP cluster damping and frequency range over $[0.04,0.25]$ and $[0.22,0.55]$. On $40\le u\le110$, the resonant potential gives normalized squared residuals $2.9281\times10^{-5}$ for a single-exponential model and $3.9822\times10^{-10}$ for a freely centered EP, both with the same ordinary response and tail terms.

Figure~\ref{fig:s_gravitational}(b) shows dependence on the starting time. For the split potential, fixing the EP center to the independently determined mean complex frequency increases its distance from the waveform. In the comparison of shared and separately fitted centers, the ordinary frequency is held at the shared optimum in both fits. The full window, response-bank, and boundary-state results accompany the numerical data.

\section{Higher-order coherent multiplets}
The split-pair response with weak relative Jordan excitation is the simplest member of a broader hierarchy. Consider $h_\delta=Ce^{s\tau}[\cosh(\delta\tau)+\xi\sinh(\delta\tau)]$ with $\xi\ne\pm1$, and set $b_\xi=1-\xi^2$, $\nu=s+\xi\delta$. The shift to $\nu$ absorbs the term linear in $\delta\tau$, giving
\begin{equation}
    \frac{h_\delta}{Ce^{\nu\tau}}
    =1+\tfrac12b_\xi\delta^2\tau^2-\tfrac13\xi b_\xi\delta^3\tau^3+O(\delta^4).
\end{equation}
The nearby EP form $C(1+u)e^{(\nu+v)\tau}e^{-z\tau}(1+z\tau)$ matches the quadratic term at $z=r\delta+O(\delta^2)$, $r^2=\xi^2-1$. If $Q_j$ projects out $g_n=\tau^ne^{s\tau}$ for $0\le n\le j$ and any fixed ordinary-response basis functions, then
\begin{equation}
    \begin{aligned}
        d_1^2&=\frac{|C|^2|b_\xi|^2}{4}\norm{Q_1g_2}^2|\delta|^4+o(|\delta|^4),\\
        d_\EP^2&=\frac{|C|^2|b_\xi|^2}{9}\norm{Q_2g_3}^2|\delta|^6
        \min_{r^2=\xi^2-1}|r-\xi|^2+o(|\delta|^6).
    \end{aligned}
\end{equation}
Both coefficients are positive. For $s=-1$, $\delta=\ii\epsilon$, and $\xi=1/2$, they are each $9/256$ after division by $E_0=|C|^2/2$.

For $K\ge3$ simple poles $s+\delta\kappa_k$ with distinct offsets $\kappa_k$, let $P(z)=\prod_k(z-\kappa_k)$ and $\ell_k(z)$ be their Lagrange interpolation polynomials. Choose $C\ne0$ and $d$ with $P(d)P'(d)\ne0$, and set
\begin{equation}
    A_k=C\ell_k(d)+j\ell_k'(d),\qquad
    j=-C\frac{P(d)}{P'(d)}.
\end{equation}
Then $\sum_kA_k\kappa_k^n=Cd^n+jn d^{n-1}$ for $0\le n\le K$. The first unmatched moment is $-jP(d)$ at $n=K+1$, giving
\begin{equation}
    d_\EP^2=\frac{|jP(d)|^2}{[(K+1)!]^2}
    \norm{Q_2g_{K+1}}^2|\delta|^{2K+2}+o(|\delta|^{2K+2}).
\end{equation}
The candidate is $(C+j\delta\tau)e^{(s+d\delta)\tau}$. Its three complex parameters remove the remaining components along $g_0,g_1,g_2$. The other quadratic-matching center has a nonzero cubic mismatch, so this branch gives agreement to higher order.

For a specific triplet,
\begin{equation}
    h=e^{s\tau}\left[-\tfrac98+\tfrac94e^{\delta\tau}
    -\tfrac18e^{2\delta\tau}\right],\qquad
    \frac{d_\EP^2}{E_0}\sim\frac{1377}{65536}\epsilon^8
\end{equation}
at $s=-1$, $\delta=\ii\epsilon$. Reversing the residues and combining the two records with amplitudes divided by $\sqrt2$ preserves total $E_0$ but gives $d_{\EP,\rm joint}^2/E_0\sim25\epsilon^4/256$. The record polynomials $\sum_kA_{jk}(\kappa_k-d)^2$ have no common root. More generally their lack of a common root is the condition for fourth-order joint separation of a fixed-residue multiplet. For more than two poles, nonproportionality alone need not imply this condition.

For two Gaussian data models with identical covariance $\sigma^2I$, the total variation distance, which determines their optimal statistical distinguishability, is $2\Phi(d/(2\sigma))-1$. Therefore $d_\EP/\sigma\to0$ makes the closest EP statistically indistinguishable from the target. For the generic split pair there is a range $\epsilon^3\ll\sigma\ll\epsilon^2$ in which departure from a simple exponential can be detected while the EP comparison remains unresolved. A coherent triplet extends the lower scale to $\epsilon^4$. The hierarchy changes the information requirement for identifying spectral structure.

\section{Kerr spectral reference}
The massive-scalar Kerr EP provides a complementary spectral example \cite{Cavalcante2024}. For separated dependence $e^{-\ii\omega t+\ii m\varphi}S(\theta)R(r)$, set $\Delta=r^2-2Mr+a^2$ and $K=(r^2+a^2)\omega-am$. The equations are
\begin{equation}
    \begin{aligned}
        0={}&\frac1{\sin\theta}\partial_\theta(\sin\theta\,\partial_\theta S)\\
        &+\left[a^2(\omega^2-\mu^2)\cos^2\theta-\frac{m^2}{\sin^2\theta}
        +A_{\ell m}\right]S,\\
        0={}&\partial_r(\Delta\partial_rR)\\
        &+\left[\frac{K^2}{\Delta}-\mu^2r^2-a^2\omega^2+2am\omega-A_{\ell m}\right]R.
    \end{aligned}
\end{equation}
Regular angular solutions and ingoing-horizon, outgoing-infinity radial solutions define the continued-fraction condition $F(\omega,a,\mu)=0$ \cite{Leaver1985}. Solving $F=\partial_\omega F=0$ on the $\ell=m=1$ branch, including the frequency dependence of the angular eigenvalue, gives
\begin{equation}
    \begin{aligned}
        a_\EP/M&=0.9994659798621,\\
        M\mu_\EP&=0.370498139508,\\
        M\omega_\EP&=0.505829799259-0.0231012944941\ii.
    \end{aligned}
\end{equation}
These values use truncation order 1600; order 1000 changes $M\mu$ by $9.8\times10^{-11}$ and $M\omega$ by $2.7\times10^{-11}$ in magnitude. The positive-phase convention of the waveform fit is $s=(-\ii\omega)^*$.

\begin{figure*}[t]
    \centering
    \subfloat[]{\includegraphics[width=.42\textwidth]{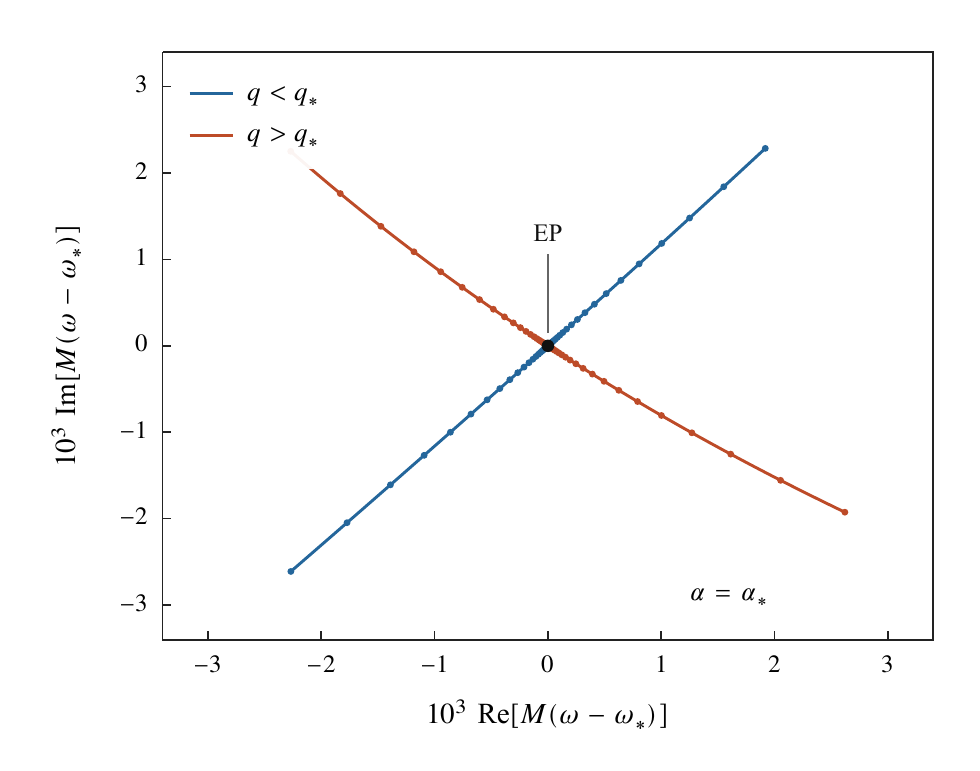}}
    \hfill
    \subfloat[]{\includegraphics[width=.42\textwidth]{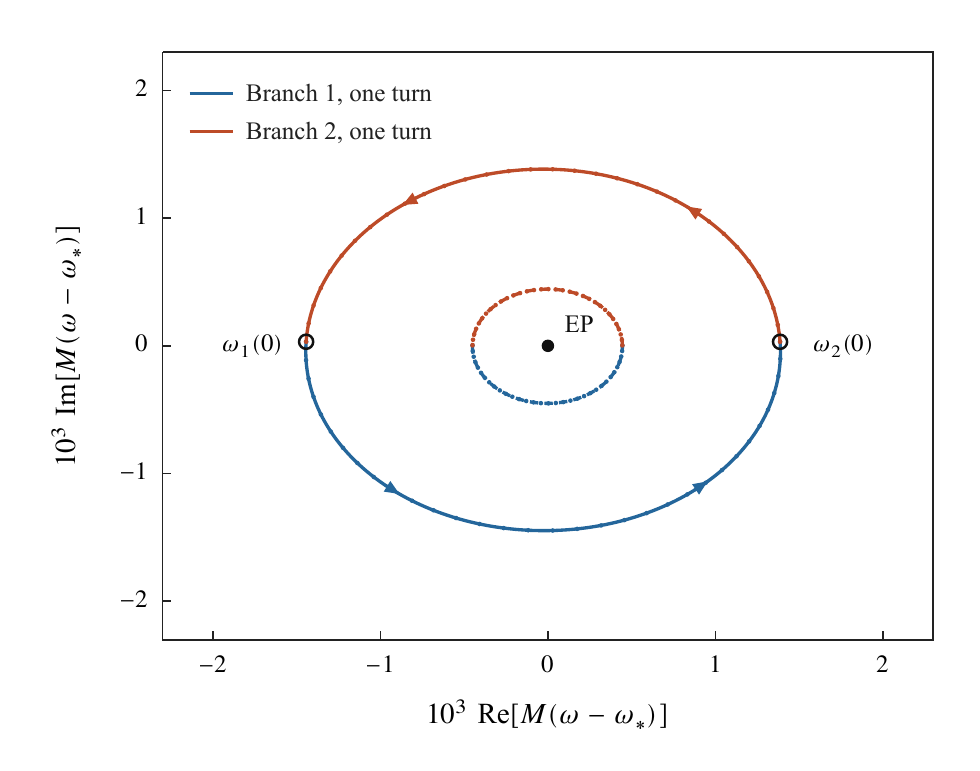}}
    \caption{Direct EMS spectra near the EP. (a) A scan at $\alpha=\alpha_*$ with 322 charge values, geometrically concentrated toward $q_*$ over $2\times10^{-9}\le|q-q_*|\le2\times10^{-5}$. (b) Continuously tracked modes during one parameter turn, $0\le\theta\le2\pi$, for $R=2\times10^{-6}$ (solid) and $2\times10^{-7}$ (dashed). Each complete two-turn calculation contains 513 parameter points. Arrows show the direction of mode tracking; the marked starting frequencies are exchanged after one turn. The plotted spectra use order 100; comparisons with orders 80 and 120 differ by at most $3.9\times10^{-9}$ along the sampled loop points.}
    \label{fig:ems_spectral}
\end{figure*}

\begin{figure*}[t]
    \centering
    \subfloat[]{\includegraphics[width=.42\textwidth]{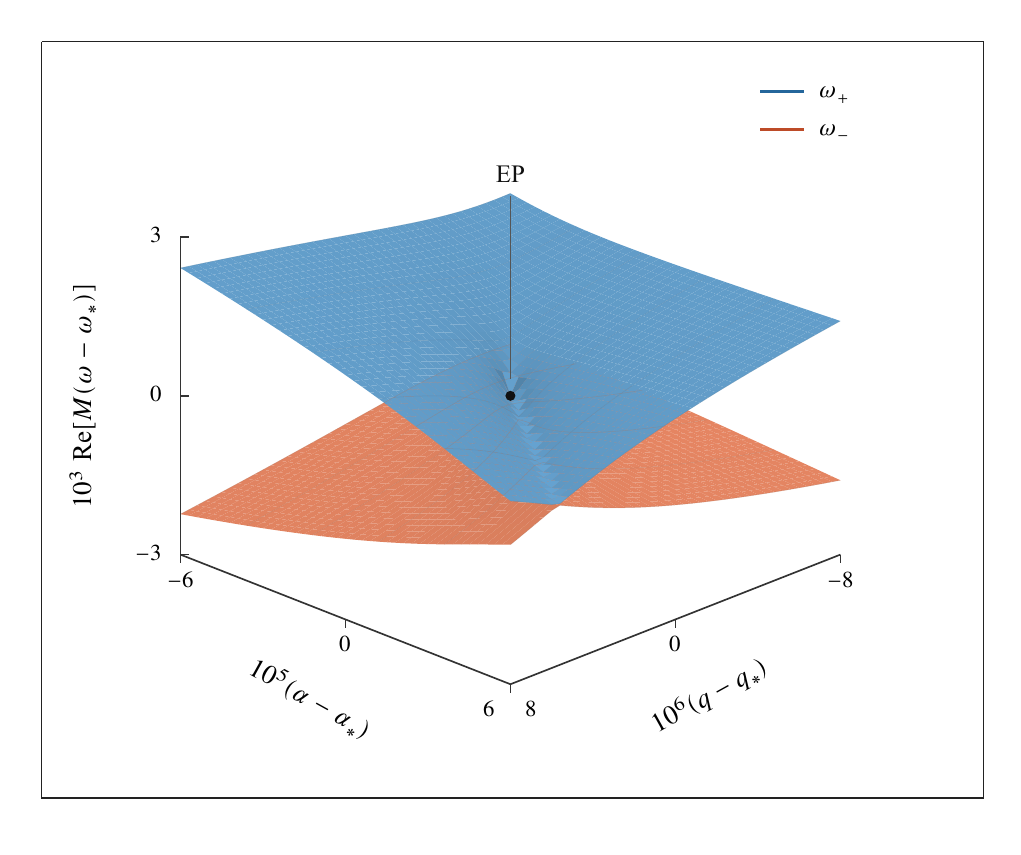}}
    \hfill
    \subfloat[]{\includegraphics[width=.42\textwidth]{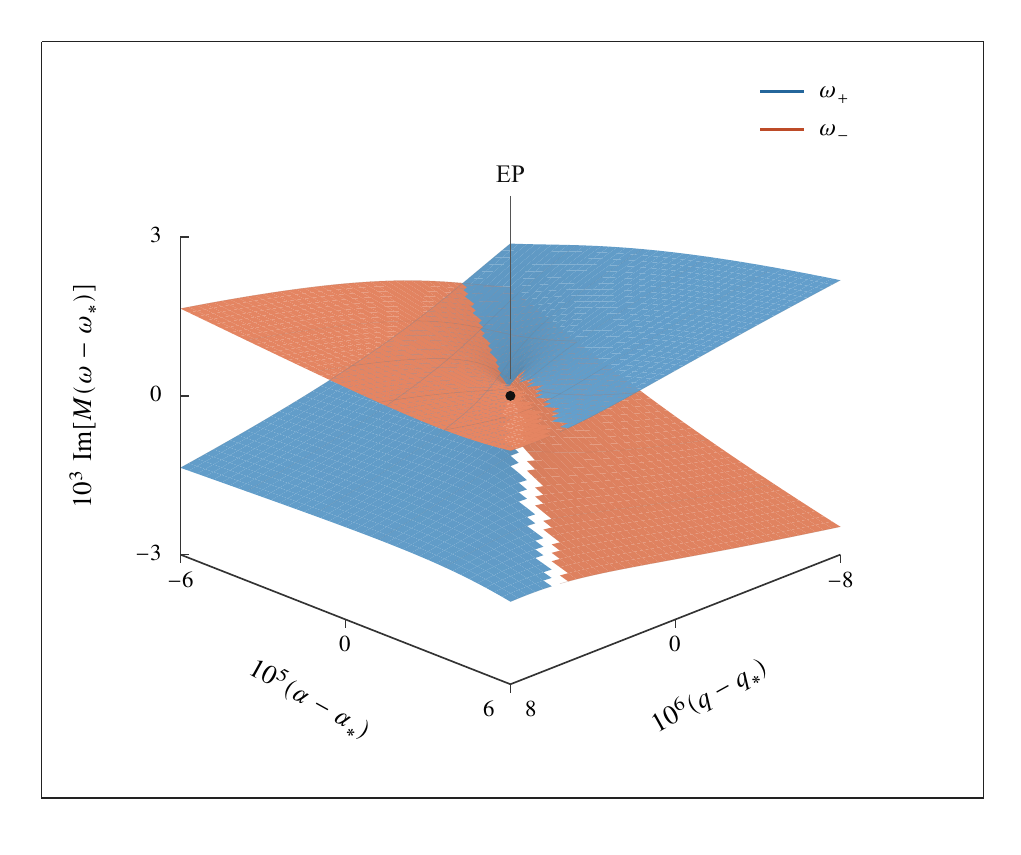}}
    \caption{Local EMS frequency surfaces from a $41\times41$ grid of independently computed spectra in the real $(\alpha,q)$ plane. (a) Real and (b) imaginary parts of the frequency displacement from Eq.~\eqref{eq:ems_spectral_ep}. The branches are labeled by their real parts, $\Ree\omega_+\ge\Ree\omega_-$; faces crossing the corresponding cut in the imaginary part are omitted. Both projections use the same computed frequency pairs.}
    \label{fig:ems_riemann}
\end{figure*}

\section{Spectral encircling of the EMS exceptional point}
\label{sec:ems_spectral}
We calculate the $\ell=2$ test-scalar spectrum directly on the EMS backgrounds near the point reported in Ref.~\cite{Cheng2026}. With $r_H=1$, we compactify the exterior by $y=1/r$ and factor the ingoing and outgoing asymptotics as
\begin{equation}
    \begin{aligned}
        \psi(y)&=e^{\ii\omega/y}y^{-2\ii M\omega}
        (1-y)^{-\ii\omega/f_H}u(y),\\
        f_H&=e^{-\delta_H}N'(r_H).
    \end{aligned}
\end{equation}
Here $\delta_H=d(r_H)$ in the metric convention of Sec.~VI. Chebyshev collocation of the regular equation for $u$ gives a quadratic eigenvalue problem in $\omega$. Both the background and the spectrum are recomputed at each parameter point. Local spectral refinement gives
\begin{equation}
    \begin{aligned}
        \alpha_*&\simeq0.79040752752,\qquad q_*\simeq1.04219114847,\\
        M\omega_*&\simeq0.65296293962-0.09023730184\ii.
    \end{aligned}
    \label{eq:ems_spectral_ep}
\end{equation}
Increasing the collocation order from $100$ to $120$ changes these coordinates and frequency components by less than $3\times10^{-12}$. The shifts from the reference parameters used in the waveform evolutions are $8.0\times10^{-9}$ in $\alpha$ and $4.7\times10^{-10}$ in $q$.

Figure~\ref{fig:ems_spectral} resolves the local spectrum and tracks both mode frequencies around the EP. We define $\mathcal D=[M(\omega_+-\omega_-)/2]^2$ and use the elliptical parameter loops
\begin{equation}
    \begin{gathered}
        \begin{pmatrix}\alpha-\alpha_*\\q-q_*\end{pmatrix}
            =R\mathsf J^{-1}\begin{pmatrix}\cos\theta\\\sin\theta\end{pmatrix},\\
            \mathsf J=\begin{pmatrix}
            -0.0418109&0.0806900\\
            0.0485211&-0.5114206
        \end{pmatrix}.
    \end{gathered}
\end{equation}
where $\mathsf J$ is the local Jacobian of $(\Ree\mathcal D,\Imm\mathcal D)$ with respect to $(\alpha,q)$. For both loop sizes, one turn exchanges the two modes and a second turn returns them to their starting values. The discriminant winds once around the origin per parameter turn, so its winding number is one. The two frequency surfaces in Fig.~\ref{fig:ems_riemann} meet at the same point, displaying the square-root branch structure of the EMS EP.

\end{document}